\documentstyle[preprint,aps]{revtex}

\def\beq{\begin{equation}}
\def\eeq{\end{equation}}
\def\beqa{\begin{eqnarray}}
\def\eeqa{\end{eqnarray}}
\def\nn{\nonumber}
\begin{document}
\begin{titlepage}
\begin{flushright}
IC/97/1\\
SISSAREF-3/97/EP
\end{flushright}
\begin{center}
{\Huge Axion production from gravitons off interacting 0-branes}

\vspace*{1cm}

{\large Faheem Hussain$^1$, Roberto Iengo$^2$ and Carmen N\'u\~nez$^3$}
\vspace*{0.5cm}

$^1$ International Centre for Theoretical Physics, Trieste, Italy\\
$^2$ International School for Advanced Studies, Trieste, Italy\\
$^3$ Instituto de Astronom\'{\i}a y F\'{\i}sica del Espacio (CONICET), 
Buenos Aires,
Argentina

\vspace*{1cm}

\end{center}

\begin{abstract}
We study axion-graviton scattering from a system of two D$0$-branes
in a Type II superstring theory, a process which does not occur 
on a single brane.  The two D$0$-branes interact via the
exchange of closed string states which form a cylinder joining them. 
By compactifying on the $Z_3$ orbifold we find 
a non vanishing amplitude coming from the odd spin structure sector,
thus from the exchanged RR states.
We compute, in particular, the leading term of the amplitude at large
distance from the branes, which corresponds to taking a field theory
limit. This seems to suggest that the process takes place through the 
coupling of an axion to the RR states exchanged between the 
0-branes.
 \end{abstract}

\begin{flushleft}
PACS: 11.25.-w\\
Keywords: string theory, D-branes
\end{flushleft}
\end{titlepage}

\section{Introduction}\label{intro}

String duality, a symmetry of string theory, gives information about the
behavior of string theory at strong coupling. It provides evidence about
 nonperturbative aspects of the theory.
This symmetry implies that 
the strongly coupled
limit of every string theory is equivalent to the weakly coupled limit
of some other theory. 
It is now believed that the known superstring theories are
different realisations of some more fundamental underlying theory.
In order to understand the nature of this unknown theory it seems necessary
 to gather as much data about it as possible by studying examples
of stringy vacua.

 Most non-perturbative string dualities require the existence of 
elementary Ramond-Ramond (RR) charges which couple to the RR fields 
contained in the spectrum of type-I and type-II superstring theories
\cite{ht,w} . As is well known string perturbation theory contains 
no such elementary RR charges. Since a $(p+1)$ form couples naturally to 
a $p$-brane, an extended object with $p$ spatial dimensions, duality 
requires us to consider such branes carrying RR charges. Such $p$- 
branes were originally found and studied as soliton solutions of the 
effective low-energy supergravity \cite{mjd}. However, following 
Polchinski's work \cite{pol1} it has become clear that there is a much 
simpler description of $p$-branes carrying RR charges. This description 
amounts to considering $p$-branes to which end points of open strings can 
be attached. This is accomplished by imposing Dirichlet boundary 
conditions on the world sheet fields, hence the name D$p$-branes or 
D-branes.

Since string theory contains gravity, the D-branes are also a source for the
gravitational field and therefore they represent a class of ``black''
objects of the family of generalized black holes. Actually, the RR charged
solitons required by string duality first appeared as black $p$-brane
solutions to the low energy effective theory. 
The conformal field theory description of D-branes  proposed in
\cite{pol1} 
has  allowed considerable progress in accounting for the 
 black hole information paradox and in probing the nature of
spacetime at the shortest distance scales.
 Examples of black holes in four and
five dimensions have been constructed for which the degeneracy of microscopic
D-brane states matches the Bekenstein-Hawking entropy \cite{stro}.
This conformal field theory formulation
consists of type II superstrings  with mixed Dirichlet
and Neumann boundary conditions. Such a world sheet approach
has made it possible to explicitly compute several properties which
had been anticipated by arguments of duality and supersymmetry. 
One of the most intriguing hints from duality is the appearance of
new length scales in string theory \cite{shenker}. By studying zero-brane
dynamics it was possible to probe distances much shorter than the
string scale \cite{kp}.
Scattering of
massless closed string 
states off D-branes at leading order in perturbation theory have
made manifest
the stringy features of the RR  solitons \cite{kt,gkm}, such as 
 the Regge behaviour and the exponential fall off of fixed angle
scattering \cite{barbon}. The exchange of massless states on the disk 
at zero momentum transfer has been
shown to agree with the $p$-brane solution to the
effective field theory at low energies. 
A systematic approach to all massless two point
functions on a disk was recently developed in \cite{mg,hk} where it was shown
that there is a direct relation between four point functions of type I theory 
and two point functions of type II theory in a D-brane background.

In this paper we study the axion production amplitude from an incoming 
graviton off a system of two D$0$-branes, or D-particles,
in a Superstring Type II theory. Actually, we will see that both choices
of Neumann or Dirichlet boundary conditions in the compactified
coordinates give the same result, and thus we speak of 0-branes 
referring to the uncompactified coordinates.   

The amplitude for this process vanishes
in the case of one D-brane \cite{mg}. Therefore, the brane-brane interaction
plays a crucial role in our interesting non trivial result
and the process can be regarded as a mechanism to
obtain novel features of the physics of D-branes.
The 
two $0$-branes interact via the exchange of closed string states which 
form a cylinder joining them. Computing the above amplitude 
implies the insertion of two NS-NS vertices on the cylinder
with the appropriate boundary conditions. It is easy to 
show that this leads to a vanishing result in 10 dimensions because of 
the abstruse identity. Having in mind a more interesting scenario, we 
consider a realistic compactification scheme breaking the 
supersymmetry down to $N=2$ in 4 dimensions. The presence of the branes 
will further reduce the supersymmetry to $N=1$. In order to be able to do 
an explicit computation we consider compactification on the standard 
$Z_3$ orbifold. This breaks enough 
supersymmetry to allow a non-zero result. 

We find that the only non-zero 
contribution to the particular axion-graviton amplitude under consideration,
 comes from the odd spin structure sector  (recall that the boundary 
conditions for world sheet fermions can be classified according to the 
spin-structure). Interestingly enough, 
the odd spin-structure corresponds to one 
term of the GSO projection of the RR world sheet fermions. 
We also find that the only 
contribution to the odd spin structure amplitude from the compactified 
coordinates arises in the twisted sector of the closed string (recall that
in an orbifold compactification twisted and untwisted sectors, by an element
of the symmetry group, have to be considered).

The branes cannot transfer energy but they can transfer momenta.
Poles in the momentum transfer arise when the vertices
corresponding to the graviton and the axion come together on
the cylinder. We refer to this process as the pinching limit. 
This pole signals the propagation of a massless closed string state,
which couples to the the point where the axion and graviton vertices
come together. It could be interpreted as a virtual axion,
propagating out of the two branes' system,  which is
eventually made real by absorbing the incoming graviton.
The residue of the pole can also be singular rather than constant.
This further singularity signals the propagation of massless 
closed string states between the branes, 
whose proper time is the length of the cylinder. The region in which
the length of the cylinder diverges can be viewed as an infrared limit of
long time propagation, the field theory limit. It corresponds to the 
exchange of the lowest closed string states, showing the suppression
of massive closed string states at large distances.
We study the amplitude in the pinching limit and field theory 
limit, taking together all the possible sources of singularities
in order to find the leading behavior for small momentum transfer,
corresponding to large distances.
The structure of the resulting expression
in this double limit seems to suggest that the process takes  
place through the exchange of an axion which couples to the RR states 
being exchanged between the D-particles.

We can also think of our amplitude containing two $0$-branes as the 
second order perturbative term in the classical (i.e. tree level) 
evaluation of the graviton-axion scattering off a single $0$-brane. 
Namely, in this picture the $0$-brane plays the role of a source, and the 
classical perturbative expansion of a nonlinear field theory  with a 
source gives rise to terms in each of which the source appears repeatedly 
\cite{duff}. We will be interested in the evaluation of the amplitude for 
small momentum transfer, that is at large distances from the source. Thus 
we could interpret the intermediate states coming out of the branes 
as the halo of the fields surrounding the gravitational object 
represented by the $0$-brane. In particular, in this paper we compute a 
scattering amplitude which is zero on one $0$-brane \cite{mg}, that 
is at 
the lowest classical perturbative order, and thus we can interpret 
the result as 
an effect due to the halo. Our computation gives the exact leading result 
for large distances.

We have organized the paper such that the arguments can be followed 
without being distracted by too much side information or computations 
and therefore 
we have kept in the main text only what is strictly necessary for
its understanding. Additional information and  
technical details have been put in appendices. In Section \ref{sec-cor} 
we compute the correlators of bosonic and fermionic world sheet fields on 
the cylinder. In Section \ref{sec-amp} we present some general 
considerations on how to construct the scattering amplitude for two 
NS-NS fields using the closed string formulation. In this section we also 
show how the ghosts are consistently taken care of, so that we can 
subsequently work in the light cone formalism. In Section 
\ref{sec-boundary} we review the construction for the RR boundary state 
and show that the compactified untwisted sector of the closed string does 
not contribute to 
the amplitude. In Section \ref{2} we calculate the final form of the 
scattering amplitude for the graviton-axion off the two $0$-branes. In 
Section \ref{pinch} we consider the pinching limit and the 
behaviour of the amplitude at small $q^2$ and discuss the result.
The content of the Appendices is the following: in Appendix A and B we
describe
the vacuum amplitude (that is the case without vertex insertions) for
the two interacting branes, first reviewing the N=4 supersymmetric
case in App.A and then constructing the orbifold case giving N=1 
in App.B. In Appendix C we show the vanishing of the even spin structures'
contribution, and in Appendix D we complete the analysis of the field
theory limit by computing subleading terms.

\section{Correlators on the cylinder}
\label{sec-cor}
\renewcommand{\theequation}{2.\arabic{equation}}
\setcounter{equation}{0}
In this section we review the construction of the cylinder and the correlators
of bosonic and fermionic fields on the cylinder.

D-branes are described by open superstrings with their end points fixed 
on a $(p+1)$ dimensional hypersurface 
embedded in ten dimensional spacetime.  The coordinates
$X^A$, ($A = 0,...,p$) with conventional Neumann boundary conditions,
span the world volume swept by the $p$-brane, whereas $X^i$, ($i=p+1,...,9$)
label the transverse directions with Dirichlet boundary conditions.

Let us consider a cylinder
joining two branes at $X^i(\sigma_1=0,\sigma_2)=0$
and $X^i~(\sigma_1~=~l,~\sigma_2~)~=~Y^i$,  
 as shown in Figure 1.  
This can be interpreted as a closed string state appearing from one D-brane,
propagating in Euclidean time $i\sigma_1 = l$ and disappearing in 
the other \cite{pol1}. $\sigma_2$ is a periodic coordinate running from $0$
to $1$. 

\medskip
\input epsf
\epsfxsize=10cm
\centerline{\epsffile{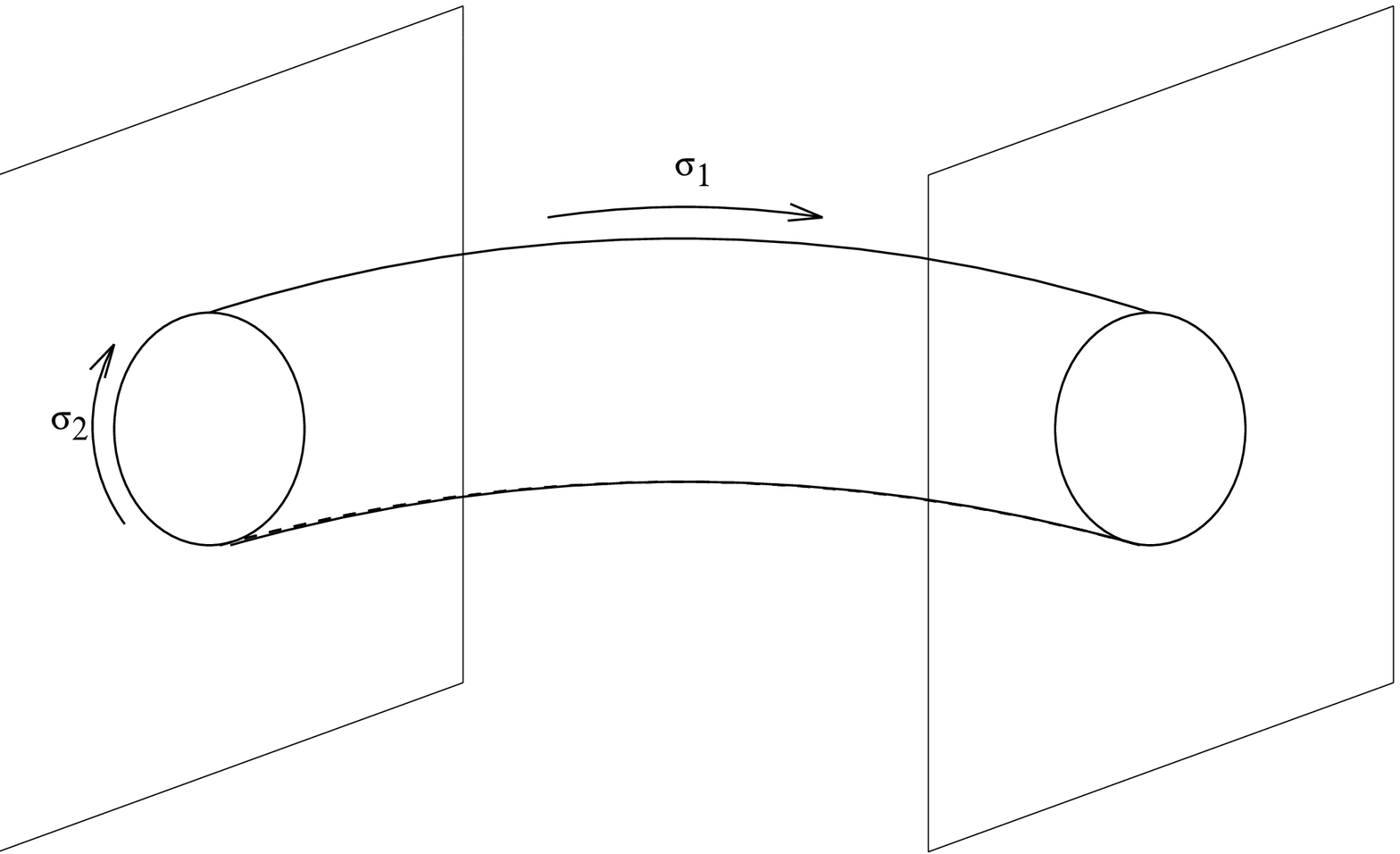}}

\medskip
{\centerline {Figure 1}}

We map the cylinder joining the two branes into a world sheet cylinder.
Following Burgess and Morris \cite{bm}
we construct this cylinder from the torus.  Using the complex coordinate
$w=\sigma_2+i\sigma_1$, the torus in Figure 2 is the rectangle 
$0\le \sigma_1\le
Im\tau$, $0\le \sigma_2 \le 1$. The cylinder is obtained by reflecting about
the dotted line $\sigma_1=Im\tau/2$ and identifying the segments OA with BC.
This corresponds to the involution $w\rightarrow\bar w+i\tau$ and leaves
invariant $\sigma_1=0$ and $\sigma_1=\tau/2$. Note that we have taken 
$\tau=i\nu$ pure imaginary, and
 $\nu= 2l$ is the real Teichmuller parameter in terms of the length of the
cylinder.

In these coordinates the Neumann boundary conditions for the bosonic
fields are 
\beq
\partial_{\sigma_1} X^A(\sigma_1=0,\sigma_2) = 
\partial_{\sigma_1} X^A(\sigma_1=l,\sigma_2) =0\,,
\eeq
whereas the Dirichlet boundary conditions, $X^{i}(\sigma_{1}=0,
\sigma^{2})=0,X^{i}(\sigma_{1}=l,\sigma^{2})=Y^{i}$ are implemented by writing
\beq
X^{i}(w)=\frac{Y^{i}Im\,w}{l}+X^{i}_{quantum}(w),
\eeq
with
\beq
X^i_{quantum}(\sigma_1=0)=X^i_{quantum}(\sigma_1=l)=0\,,
\eeq
and $Y^{i}$ is the separation of the D-branes in the $i$ direction.

\bigskip
\input epsf
\epsfxsize=10cm
\centerline{\epsffile{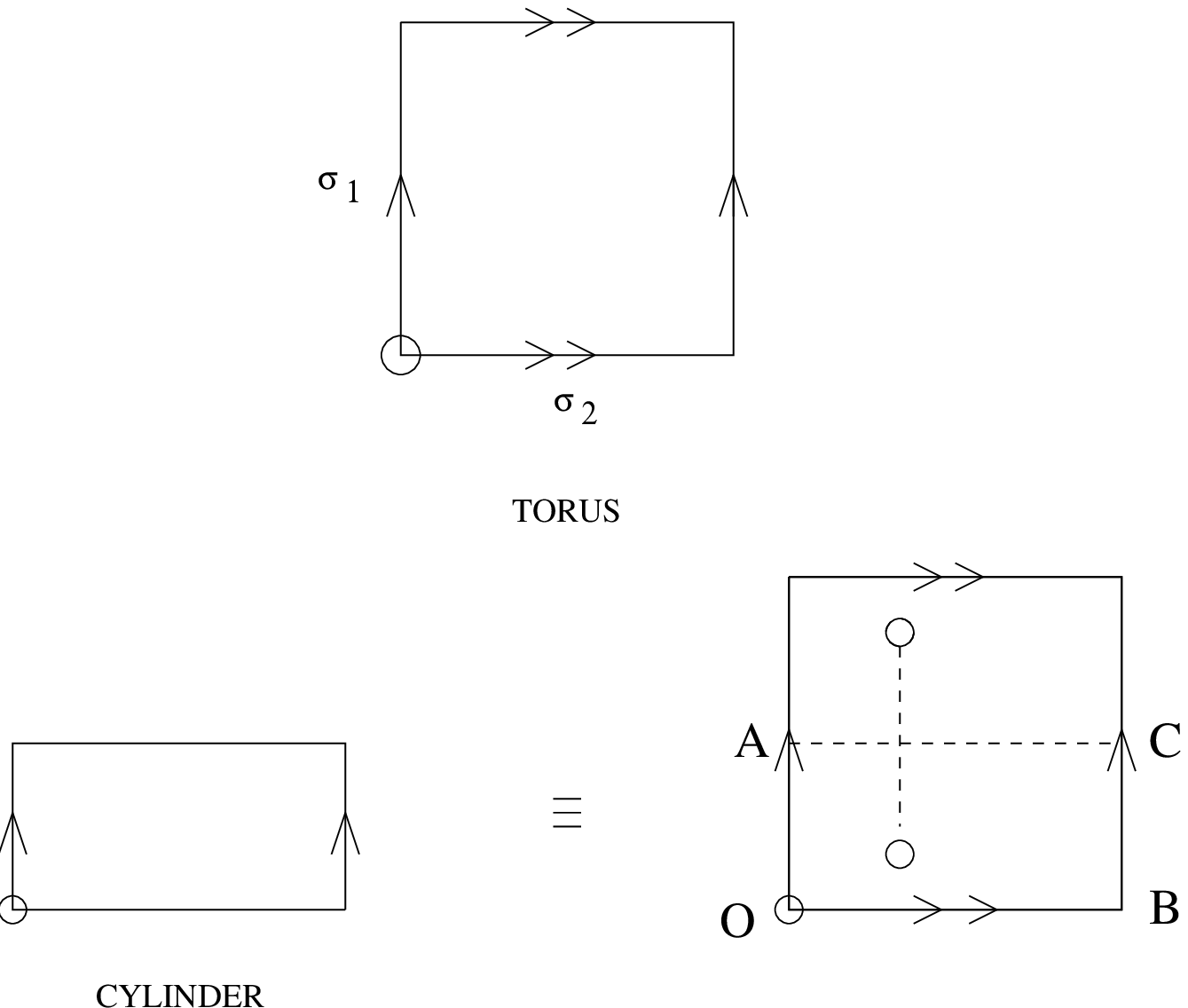}}

\bigskip
{\centerline {Figure 2}}

With these boundary conditions, we can then construct the bosonic and
fermionic propagators on the cylinder using the usual torus propagators.
For fermions, the operator representing the involution $\sigma_1\rightarrow
1-\sigma_1$ is $\gamma_1$ \cite{bm} in the two dimensional spinor space. 
It follows from world sheet supersymmetry that the Neumann boundary 
conditions are $\Psi^{A}(\bar w+2il)=\gamma_1\Psi^{A}(w)$, where 
$\Psi ={\psi(w) \brack \bar{\psi}(\bar{w})}$, giving 
$\bar{\psi}^{A}(\bar w)=\psi^{A}(\bar w+2il)$, whereas for Dirichlet
boundary conditions we get $\bar\psi^i(\bar w)=-\psi^{i}(\bar w+2il)$. 
Here and subsequently $\psi\,\,(\bar{\psi})$ mean right (left) moving 
fields respectively.
We can drop the shift $2il$ from the argument, which just brings us back
to the upper half plane into the torus. In other words instead of working in
the upper half plane we can equally well work in the strip $-il<Im\,z<il, 
0<Re\, z<1$. Also in
this paper we are only concerned with  D$0$-branes in the uncompactified
directions. The vertices that we need to insert on the cylinder
contain only uncompactified string coordinates.
Thus we can restrict here the discussion to the case in which the
coordinate index is $\mu=0, 1, 2, 3$. The coordinates and the fermions in 
the time direction have Neumann boundary conditions, whereas the ones in 
the three space directions have Dirichlet conditions. Thus the 
boundary conditions on the fermions can be written in a compact form as
\beq
\bar{\psi}^{\mu}(\bar{z})=S^{\mu}_{\nu}\psi^{\nu}(x),\label{bcf}
\eeq
with $x=\bar{z}$ and $S^{m}_{n}=-\delta^{m}_{n},\,\, S^{0}_{0}=1$ and off 
diagonal $=0$.
Thus we only need to specify the $\langle \psi(x_1)\psi(x_2) \rangle$ 
correlator, $x_1,x_2$ being complex coordinates.

The above notation can be used for the correlators of the bosonic coordinates,
which can be written as
\beq 
<X^{\mu}(z,\bar{z})X^{\nu}(w,\bar{w})>=-\eta^{\mu\nu}G(z,w)
-S^{\mu\nu}G(z,\bar{w})
\eeq
where
\beq
G(x_{1},x_{2}) = ln \left |{{\vartheta_1(x_{1}-x_{2}|\tau)}\over
{\vartheta'_1(0)}}\right |^2 - 2\pi {{(Im(x_{1}-x_{2}))^2}\over Im\tau}\label{G} 
\eeq
and $\vartheta_1(x_{1}-x_{2}|\tau)=
\vartheta{1/2 \brack 1/2 }(x_{1}-x_{2}|\tau)$.

For the fermionic correlators one has to specify the spin structure
$s$.
For the three even spin structures we have
\beqa
<\psi^\mu(x_{1})\psi^\nu(x_{2})>_{s} 
= -\eta^{\mu\nu} {\vartheta_s(x_{1}-x_{2}|\tau)\over
\vartheta_1(x_{1}-x_{2}|\tau)} {\vartheta'_1(0)\over\vartheta_{s}(0)}
 = -\eta^{\mu\nu} F_s(x_{1}-x_{2}),\label{eferprop}
\eeqa
where $\vartheta_s(x_{1}-x_{2}|\tau),\,\,s=2,3,4$ are defined in
Appendix A.
As for the $<\psi\psi>_{odd}$ correlators, they are determined by the
requirement that they have one pole at $1/(x_{1}-x_{2})$, like 
$\partial_{x_{1}}G(x_{1},x_{2})$. In fact, since in the odd spin
structure case $\psi$ has the same periodicity conditions as $X$, these
requirements give 
\beqa
<\psi^\mu(z) \psi^\nu(w)>_{odd} 
= -\eta^{\mu\nu} \partial_{x_{1}}G(x_{1},x_{2})  
 = -\eta^{\mu\nu} F(x_{1}-x_{2}).\label{ferprop}
\eeqa
Remember that, due to eq.(\ref{bcf}), eqs.(\ref{eferprop}) and (\ref{ferprop})
also encode the $<\psi \bar\psi>$ and $<\bar\psi \bar\psi>$ propagators.
Actually the odd case requires a more complete discussion. We come back 
to that at the end of the next section.

\section{Construction of scattering 
amplitude}\label{sec-amp}
\renewcommand{\theequation}{3.\arabic{equation}}
\setcounter{equation}{0}
As mentioned in the introduction we consider the scattering of NS-NS
fields from a system of two $0$-branes in a superstring theory
compactified down to four dimensions on the standard $Z_3$ orbifold.
The cylinder depicted in Figure 1 is a one loop open string graph.
The world line of the open string boundary can be regarded as a state
connecting the vacuum to one closed string.
Thus one has to first construct boundary states as we do our
calculation in the closed string formulation. Our
$0$-branes are such that only the time coordinate $X^{0}$ has Neumann
boundary conditions. The three uncompactified space coordinates have 
Dirichlet boundary conditions. Using the conventions of ref. \cite{pol2}, 
the Neumann and Dirichlet boundary conditions
translate into the following conditions for the boundary state ($n$ means
modes of the $\sigma_{2}$ Fourier expansion):
\beq
  (\alpha_n^0 + \tilde{\alpha}_{-n}^0) |B> = 0,\quad
(\psi_n^0 
+ i\eta \tilde{\psi}^0_{-n} ) |B> = 0\label{bon}
\eeq
and
\beq
  (\alpha_n^i - \tilde{\alpha}_{-n}^i) |B> = 0,\quad
(\psi_n^i 
- i\eta \tilde{\psi}^i_{-n} ) |B> = 0\label{dbc}
\eeq
for all $n$. Here $i=1,2,3$ denote the three uncompactified space
directions and  the tilded or
untilded operators correspond to the mode expansion of right and left
movers. Further,  $\eta=\pm 1$ and we will see that the different
spin structures arise from taking the same or opposite value of
$\eta$ for the boundary states of the two branes respectively.

Here we have lumped together NS-NS and R-R sectors. These are specified
by taking $n$ to be half-integer or integer, repectively
in the mode expansion for the fermions. To be 
general we also have to specify the boundary state conditions for the six
compact coordinates, $X^{4}\cdots X^{9}$. One has to keep several 
things in mind here. The boundary state condition will include not only 
the spin structure factor $\eta$ and Neumann
(Dirichlet) signs but one has also to project onto the $Z_{3}$ invariant
states \cite{minahan}. We  shall discuss this further later in this section. 
Also we have to remember that in the $\sigma_{2}$ twisted sectors the moding 
is not integral or half integral but in integer 
multiples of $1/3$.
These conditions implicitly define the boundary state $|B, k>$, with 
momentum $k$.

Then the boundary states on the two $0$-branes are given by
\beqa
|B, X^i = 0> & = & \int  {{d^{3}k} \over {(2\pi)^{3}}} e^{i k\cdot (X=0)} 
|B,k> \nonumber \\
|B, X^i = Y^i> & = & \int {{d^{3}k} \over {(2\pi)^{3}}} e^{i k\cdot Y} 
|B,k>
\eeqa

Now we review the general construction of the scattering amplitude,
leaving the details to be explained in other Sections.
The scattering amplitude $A$ can be written as a path integral over
world-sheet fields on a cylinder. The integrand is the 
product of two NS-NS vertices, $V$ and $U$, in the 0-ghost
 picture, constructed 
from world sheet fields with components only in the uncompactified 
directions (remember that on the cylinder the 
worldsheet left and right moving fields are identified up to some phase).
The result of the integration can be cast in the form
\beq
A(l;z,w)=\sum_s (\pm) <U(z,\bar{z})V(w,\bar{w})>_s\cdot Z_s\,,      
\label{scatamp}
\eeq
where $<UV>_s$ is evaluated in terms of correlators by using the
Wick theorem, and $Z_s$ is the socalled partition function given by
\beq
Z_{s} = <B|e^{-lH_s}|B>_{s}\,,\label{partf} 
\eeq
where $H_s$ is the relevant Hamiltonian for the world sheet fields.
For the fermionic coordinates one has to specify the 
spin structure $s$ and sum over the spin structures with appropriate 
signs as indicated above. The four spin structures correspond in the
operatorial language to the two terms of the GSO projection
for each of the NS-NS and RR fermions. 
Eventually one has to integrate over the $z,w$ vertex positions
on the cylinder and over the cylinder length $l$, giving the final 
scattering amplitude as
 \beq
{\cal M}=\int_0^\infty dl l^{-\frac{3}{2}}e^{-{Y^{2}}\over{2l}}\int\int
d^{2}zd^{2}{w}A(l;z,w)\,.\label{amp}
\eeq 
We are interested in the case when the vertices only contain
uncompactified coordinates. In principle $A(l;z,w)$ also contains a sum
over the discrete compactified momenta $p_{n}$ (they are zero if there are
Neumann boundary conditions in the corresponding directions, or for
Dirichlet compactified directions one has a wave function like for the
uncompactified case). Since later we will be interested in the
$l\rightarrow\infty$ corner of the moduli space (because it will
correspond to the field theory limit) and since each momentum $p_{n}$ is
weighted by a factor $e^{-lp_{n}^{2}}$ from the compact Hamiltonian,
only the $p_{n}=0$ term of the sum will be relevant. In conclusion, we
can take the momenta in the compact direction to be zero for both the
Neumann and Dirichlet compactified directions. Correspondingly we refer
to our branes as $0$-branes, even when they could be Neumann in the
internal compact directions.

The partition function is the result of the functional integration
without vertex insertions
 (i.e. there is 1 in the place of the vertices). It is a product of
the integration over the bosonic coordinates,  the $b-c$ ghosts,
 the fermionic coordinates and  the $\beta -\gamma$ ghosts.

The $b-c$ ghost contribution cancels the contribution of the bosonic pair
$X^0X^1$. The $\beta -\gamma$ ghost contribution is like the inverse of the
fermionic pair $\psi^0 \psi^1$ for every spin structure, and thus it
cancels this latter contribution.

Thus, the partition function has the form of the ``light-cone" expression
\beq
Z_s=Z^B_{LC}\cdot Z^F_{LC,s}\,,\label{partfunct}
\eeq
where $Z^B_{LC}=\prod_{\mu =2,4,6,8} Z^B_{\mu}$ is the product 
of the contributions of the $X^{\mu} ,X^{\mu +1}$ pairs,
and $Z^F_{LC,s}=\prod_{\mu =2,4,6,8} Z^F_{\mu,s}$ is the product 
of the contributions of the $\psi^{\mu} ,\psi^{\mu +1}$ pairs.

With our conventions 
$\{\psi^{\mu}_n ,\psi^{\nu}_{-n} \} =\eta^{\mu \nu}$, with 
$\eta^{00} =-1$, and  
$(\psi^{\mu}_n )^{\dag} =\mp \psi^{\mu}_{-n} $ with upper/lower
sign for $\mu =0$ or $\not= 0$. 
Thus $(\pm\psi^0_n + \psi^1_{-n} )/\sqrt{2}$
and, for $\mu\not= 0$, $(\psi^{\mu}_n \pm i\psi^{\mu +1}_{-n} )/\sqrt{2}$
are like fermionic destruction or creation operators.
The contribution of each pair is derived in Appendix B. 

In order to compute $<UV>_s$ we need the explicit form of the
correlators which are given in Section~\ref{sec-cor}. Correlators can 
involve a pair of $X$'s, a pair of $\psi$'s, a pair of $\bar\psi$'s, or 
one $\psi$ and one $\bar\psi$. 

We show in Appendix C that for the particular case of the
graviton-axion amplitude, which we consider in this paper, only the odd
spin structure contributes. We recall that the odd spin structure
corresponds to a part of the GSO projection of the RR fermions.
We also anticipate that the $\sigma_{2}$ untwisted case for the
compactified coordinates does not contribute for the relevant odd spin 
structure. 

In the odd spin structure case the fermionic coordinates have zero modes
and therefore the result will be zero unless the amplitude contains a
sufficient number of fermionic fields. 
In the relevant - twisted - orbifold case the zero modes only appear for the 
noncompactified coordinates $\mu =0,1,2,3$. Thus, only
the terms in the expression of $<UV>$ containing a factor
$\psi^0 \psi^1 \psi^2 \psi^3$ will contribute to the path 
integral (remember that some $\bar \psi$'s can appear in the 
place of $\psi$, with the identification of eq.(\ref{bcf})). 
We can still express the result in the form of
eq.(\ref{scatamp}), with the understanding that the partition function 
contains the zero modes insertion.
For this odd spin structure sector we obtain
 $Z^B_{LC}\cdot Z^F_{LC,odd}=1$, as in 
this case the part of the GSO projection of the RR fermions
 exactly compensates 
the bosonic contribution (see Appendix B). 

Therefore in eq.(\ref{scatamp}) we need to compute the remaining part
of $<UV>_{odd}$. Altogether this amounts to saying that we compute
$<UV>_{odd}$ by Wick's theorem, using the correlators specified above,
and factorizing in each term a factor 
\beq
<\psi^{\mu}\psi^{\nu}\psi^{\rho}\psi^{\sigma}>_{odd}
=-i\epsilon^{\mu\nu\rho\sigma}
\eeq 
to keep into account the hermiticity properties and 
the sign of the different permutations. 
We recall again that
\beq
<\psi^{\mu} (x_1) \psi^{\nu} (x_2)>_{odd}=
     -\eta^{\mu \nu} \partial_{x_1} G(x_1,x_2)
\eeq 
and note that the nonholomorphic part of the rhs actually  
cancels when summing the various terms of $<UV>_{odd}$.

\section{Boundary state for the RR case}\label{sec-boundary}
\renewcommand{\theequation}{4.\arabic{equation}}
\setcounter{equation}{0}
Here we review the construction of the boundary state
for the RR case and show that the $\sigma_2$ untwisted odd spin structure 
does not contribute to our amplitude.

Consider first the four uncompactified fermionic coordinates
$\psi^{\mu}$, with $\mu =0,1,2,3$. When performing the Fourier
analysis in $\sigma_2$ we get modes $n=0$ in the RR case:
they can be identified with the $\gamma$-matrices 
$\gamma^{\mu}=i\sqrt{2}\psi_{0}^{\mu}$ and 
$\tilde{\gamma}^{\mu}=i\sqrt{2}\bar{\psi}_{0}^{\mu}$, with 
$\{\gamma^{\mu},\gamma^{\nu}\}=-2\eta^{\mu\nu}$, which act on a subspace
which is a direct product of two spinor spaces $S_i\otimes \tilde{S}_j$.
Let us examine the sector of the
boundary state related to this subspace. Since we have Neumann 
boundary condition in time and Dirichlet in space, we have
for the boundary state $|B\eta >$ 
\beqa
(\gamma^0 +i\eta \tilde\gamma^0 )|B\eta > & = & 0 \nonumber \\
(\gamma^{\mu\not= 0} -i\eta \tilde\gamma^{\mu\not= 0} )|B\eta > & = & 0
\eeqa
where $\eta =\pm 1$ has been put for later convenience.
Let us define $a =(\gamma^0 +\gamma^1 )/2$, 
$a^*~ =~(\gamma^0~-~\gamma^1)/2$
and $b= (-i\gamma^2 +\gamma^3 )/2$, $b^*= (-i\gamma^2 -\gamma^3 )/2$
such that $\{ a,a^* \}~=~\{ b,b^*\}~=~1$, similarly for $\tilde a,\tilde b$, 
and zero for the other anticommutators.  
The conditions on $|B\eta >$ can now be rewritten as
\beqa
(a+i\eta \tilde{a}^*)|B\eta >  = 0, && ~~~
(b-i\eta \tilde{b})|B\eta > = 0 \nonumber \\
(a^* +i\eta \tilde{a})|B\eta > = 0, && ~~~~(b^*-i\eta\tilde{b}^*)|B\eta >  = 0 
\eeqa
Defining a ``vacuum" $|0>\otimes |\tilde 0>$ by
$a|0>=b|0>=0,\tilde a|0>=\tilde b^*|0>=0$, we find the boundary
state
\beq
|B\eta >={1\over \sqrt 2}
e^{-i\eta (a^*\tilde a^*-b^*\tilde b)}|0>\otimes |\tilde 0>
\eeq
which can be expressed as a sum of products of spinors
\beq
|B\eta >= \sum_{i=1}^4 c_i(\eta )S_i\otimes \tilde{S}_i
\eeq
One can construct explicitly the spinors and  
see that each $S_i, \tilde{S}_i$ is an eigenstate of
$\gamma^5, \tilde \gamma^5$ and moreover that for each $i$:
$\gamma^5 S_i=-\tilde\gamma^5 \tilde{S}_i$. 

Thus, the boundary state for the $0$-brane is only compatible with
Type IIA theory \cite{pol1}. Also, one can analyse the spacetime content 
of the boundary state by computing 
$\sum_ic_i(\eta )S_iC(\gamma~-matrices)\tilde{S}_i$,
where $C$ is the charge conjugation matrix. It is seen that
the result is different from zero only for the case of
$\gamma^0$ and $\gamma^5\gamma^0$, consistently with the
picture that the 0-brane at rest is the source for the
$0$-component of a vector or axialvector. 

Moreover
\beq
\gamma^5 |B\eta >=-\tilde \gamma^5 |B\eta >=-|B-\eta >
\eeq
The GSO projected partition function 
\footnote{Recall that 
the $n=0$ fermionic modes do not enter into the Hamiltonian which allows 
us to write the partition function, eq. (\ref{partf}), in this simple 
symbolic form where $|B\eta>$ refers only to the $n=0$ modes.}  
can be written as 
\beq {1 \over 2}(<B\eta |B\eta >-<B-\eta |B\eta >)\,.
\eeq
Due to the fact that changing the sign of the $\bar\psi$'s
in $|B\eta >$ is the same as changing the sign of $\eta$,
the GSO projection for the  $\sigma_{2}$-Fourier $n=0$ modes
amounts to the projection
\beq
<B\eta |{{1+\gamma^5}\over 2}{{1-\tilde\gamma^5}\over 2}|B\eta >\,.
\eeq
Actually $<B\eta |B\eta >=2$ and $<B-\eta |\gamma^5|B\eta>=-2$, whereas 
$<B-\eta |B\eta >=0$ which corresponds to the fact that the odd spin 
structure partition function is zero. The odd spin structure contribution
is different from zero when one inserts 
$\psi^{\mu}\psi^{\nu}\psi^{\rho}\psi^{\sigma}$ which here
is the same as $-i\epsilon^{\mu\nu\rho\sigma}\gamma^5$, 
consistently with rules for constructing the scattering
amplitude in the odd spin structure case, see Section~\ref{sec-amp}. 

Of course, in order to construct the complete boundary state
one has to include the $\sigma_2$-Fourier $n\not= 0$ part,
and also to include the compactified coordinates part.

In general, the full partition function is
\beq
Z=\frac{1}{3}\sum_{\{g\}}\sum_{\{h\}}Z(total)_{g,h}
\eeq
where $\{g\}=1,g,g^2$ and $\{h\}=1,h,h^2$ are twists along the $\sigma_1$ 
and $\sigma_2$ directions for the fields corresponding to the
compactified directions. 
Here by $Z(total)$ we mean $Z(total)~=~Z^{bosons}.Z^{fermions}$. Remember
that one has also to
project over the $Z_3$ invariant states, and this is implemented by the
average over ${g}$.
The compactified fermionic coordinates have zero modes
in the $\sigma_2$-untwisted sector only.  Now 
consider the $Z(total)$ just for the untwisted compactified sector and for 
the odd spin structure only. In this case the 
contributions to the partition function from the bosonic and fermionic 
modes for $n\not=0$ exactly compensate leaving
\beq
Z^{c}_{g,1}= Z_{g}^{c,fermions(n=0)}\,,
\eeq 
where the superscript $c$ is used to indicate the compactified part.
One can now construct the action of $g$ on the boundary state in the 
following way.
Namely, consider the (4,5) (6,7) (8,9) pairs, and construct for each of 
them the $b,b^{\dag}$ operators like we have done for the (2,3)
pair. The action of an element $g$ of $Z_3$ amounts to  
$b_{(j,j+1)} \rightarrow g_j b_{(j,j+1)}$, with 
$g_j=e^{{{2\pi iz_j}\over 3}}$ such that $z_4+z_6+z_8=0$ mod 
$2$\cite{minahan} (see Appendix B). Denoting by $|B_c\eta >$
the boundary state for the $n=0$ modes of the compactified 
sector, we thus have
\beq
g\cdot |B_c\eta >=\prod_{j=4,6,8} 
e^{i\eta g_j (b^*\tilde b)_{j,j+1}}|0>\otimes |\tilde 0>
\eeq
and the $Z_3$ invariant combination is
\beq
|B_c\eta >_{inv}=\frac{1}{3}(1+g+g^2)|B_c\eta >
\eeq
The odd spin structure contribution is found to vanish:
\beq
<B_c-\eta |B_c\eta >_{inv}=0
\eeq
because of the condition $\sum_l z_l = 0$ mod $2$.
Since our vertices do not contain compactified coordinates,
the $\sigma_2$-untwisted odd spin structure case does not
contribute. 

For us, therefore, the relevant case is the $\sigma_2$ twisted sector of the
orbifold, where there are no $n=0$ modes for the compactified
coordinates. Besides, as we have said in Section \ref{sec-amp}, the odd 
spin structure partition function with the suitable insertion 
of $i\psi^{0}\psi^{1}\psi^{2}\psi^{3}$ is equal to $1$,
since the bosonic contribution compensates the
fermionic contribution, for each higher-$n$
mode, both in the uncompactified and in the 
(twisted) compactified sector, and also for the ghosts.  
Thus, in this case we have just to compute the 
correlator $<UV>_{odd}$, with the rules of Section \ref{sec-cor}.

\section{Two point functions on the cylinder}\label{2}
\renewcommand{\theequation}{5.\arabic{equation}}
\setcounter{equation}{0}
Let us consider the scattering of two massless string states ($p^2=0$) 
whose vertex operators are

\beq
V(z,\bar z) = \epsilon_{\mu\nu} (\partial X^{\mu}(z,\bar{z}) + i p\cdot \psi(z) \psi^{\mu}(z))
(\bar\partial X^{\nu}(z,\bar{z}) 
+ i p\cdot \bar\psi(\bar{z}) \bar\psi^{\nu}(\bar{z)}) e^{ip\cdot X(z,\bar{z})}
\eeq
with $\epsilon_{\mu\nu}$ transverse ($p^{\mu}\epsilon_{\mu\nu}
= p^{\nu}\epsilon_{\mu\nu}=0$) and symmetric for gravitons 
($\epsilon_{\mu\nu}=\epsilon_{\nu\mu}; \epsilon_\mu^\mu=0$) and dilatons
($\epsilon_{\mu\nu}=\eta_{\mu\nu}-p_\mu l_\nu-l_\mu p_\nu$, 
where $p\cdot l = 1$) and antisymmetric for antisymmetric tensor particles.
Here $\partial\equiv \partial_{z}$ and 
$\bar{\partial}\equiv \partial_{\bar{z}}$.
The general expression for the scattering amplitude on the cylinder has been
given as equation~(\ref{amp}) in Section~\ref{sec-amp}.

We show in Appendix A that summing over all spin structures, amounting
to the GSO projection, 
yields a vanishing vacuum functional
at the one-loop level \cite{pol1}. Recalling the analysis in Ref.\cite{nns}
it is easy to show that one loop amplitudes for up to two external legs
also vanish in ten dimensions. 

However, when this process is considered on the orbifold
the result is more interesting.
Without loss of generality, we take the time as the unique Neumann direction,
 and 
the polarization tensors to be non-vanishing only in the directions 
perpendicular to the 0-brane, i.e. $\epsilon_{00}=\epsilon_{0i}=\epsilon_{i0}
=0; \epsilon_{ij}\ne 0$.

We now explicitly consider the scattering of a graviton and an antisymmetric
tensor, with polarization tensors $h_{ij}$ and $b_{ij}$
respectively, off the two 0-branes. The two vertex operators are,
respectively,
\beqa
V_{h}(k,z,\bar z) & = & h_{ij} (\partial X^{i}(z,\bar{z}) 
+ i k\cdot \psi(z)\psi^{i}(z))
(\bar\partial X^{j}(z,\bar{z}) 
+ i k\cdot \bar\psi(\bar{z}) \bar\psi^{j}(\bar{z})) e^{ik\cdot
X(z,\bar{z})}\nn\\
V_{b}(p,w,\bar w) & = & b_{lm} (\partial X^{l}(w,\bar{w}) 
+ i p\cdot \psi(w)\psi^{l}(w))
(\bar\partial X^{m}(w,\bar{w}) 
+ i p\cdot \bar\psi(\bar{w}) \bar\psi^{m}(\bar{w})) e^{ip\cdot
X(w,\bar{w})}\nn\\
&&
\eeqa
with
\beqa
k_{i}h_{ij}=0,\quad h_{ij}=h_{ji}\quad
\sum_{i}h_{ii}=0\nn\\
p_{l}b_{lm}=0,\quad b_{lm}=-b_{ml}.
\eeqa

We show in Appendix
C that the contribution from the even spin structures vanishes.
Let us then
consider the odd spin structure sector. 
As we have seen in Section~\ref{sec-amp} a non vanishing result requires
the presence of four fermionic zero modes. We
denote the possible zero mode contributions as
\beqa
<\psi^\mu\psi^\nu\psi^\rho\psi^\sigma> = (-i)\epsilon^{\mu\nu\rho\sigma},
\quad
<\psi^\mu\bar\psi^\nu\psi^\rho\psi^\sigma> = (-i)
\epsilon^{\mu\tilde\nu\rho\sigma}\nn\\
<\psi^\mu p\cdot\bar\psi^\nu\psi^\rho\psi^\sigma> = (-i) 
\epsilon^{\mu\nu\rho\sigma}\tilde p_\nu\,,\,etc.
\eeqa
where $\tilde{\nu}$ means that one must change the sign
whenever it is a space index. Similarly $\tilde{p}_{\nu}$ signifies that
the space part is opposite to the space part of $p_{\nu}$.

It is convenient to summarize the odd fermionic propagators,
equation~(\ref{ferprop}), once again in
this section as follows:
\beq
<\psi^{\mu}(z)\psi^{\nu}(w)> = -\eta^{\mu\nu}F(z-w),\quad
<\psi^{\mu}(z)\bar{\psi}^{\nu}(\bar{w})> =
-\eta^{\mu\tilde{\nu}}F(z-\bar{w})\,,
\eeq
where
\beqa
F(z-w) & = & \frac{\vartheta_{1}^{\prime}(z-w)}{\vartheta_{1}(z-w)} 
+\frac{i\pi}{l}(Imz-Imw)\nonumber\\
F(z-\bar{w}) & = & \frac{\vartheta_{1}^{\prime}(z-\bar{w})}{\vartheta_{1}(z-\bar{w})} 
+\frac{i\pi}{l}(Imz+Imw)\,.\label{F}
\eeqa
Another useful piece of notation is the contraction
\beq
<k\cdot\psi(z) p\cdot\bar\psi(\bar{w})> = -k\cdot \tilde p F(z-\bar{w})
\eeq
where $k\cdot \tilde p = -{\bf k}\cdot {\bf p} - k_0 p_0$.

The normal ordering of the
exponential factors gives
\beq
e^{-k_\mu p_\nu <X^\mu(z) X^\nu(w)>} \equiv
\left \vert {\vartheta_1(z-w)\over \vartheta_1(z-\bar w)} \right \vert ^{q^2}  
\left \vert {\vartheta_1^2(z-\bar w)\over{\vartheta_1(z-\bar z)\vartheta_1(w-\bar w)}}
\right \vert ^{2k_0^2}e ^{{2\pi\over l} q^2 Im z Im w +
{2\pi k_0^2\over l}(Im z- Im w)^2} \label{exp}
\eeq
after using the following kinematical relations
\beqa
p^0 = -p_0 = k_0\nn\\
q^2 = (k+p)^2 = 2k\cdot p\nn\\
k_i^2 = p_i^2 = k_0^2. \label{kin}
\eeqa
Note that the branes cannot transfer energy, but they can transfer
momenta. Thus $q^{2}$ is purely spacelike ($q^2=\vec{q}^2$).

Instead of using the antisymmetric polarization tensor, $b_{lm}$, we
will write our amplitude in terms of the axion $a$ introduced as
\beq 
-p_{0}b_{lm}=\frac{a}{2}\epsilon_{lms}p_{s}\,,
\eeq
where $\epsilon_{lms}$ is now the usual Levi-Civita symbol. 
One now does the Wick contractions, remembering that four fermions have
always to be taken as zero modes to get a non-zero answer. This means
that one only gets contributions when there are four, six or eight fermion
fields. After some lengthy calculations we find

\beqa
<V_{h}(k,z,\bar{z}) V_b(p,w,\bar{w})>& = &e^{-k_\mu p_\nu 
<X^\mu(z)X^\nu(w)>}a ~  q\cdot h\cdot q\left[\frac{q^{2}}{2}
 \left\{-|F(z-w)|^2
+ |F(z-\bar w)|^2 \right.\right.\nn\\
&&\hspace{-1.0in}\left.+ {1\over 2}(F(z-w))^2 
+ {1\over 2} (F(\bar z- \bar w))^2
- {1\over 2} (F(z-\bar w))^2 - {1\over 2} (F(\bar z-w))^2\right\}\nn\\
&&\hspace{-1.0in} - 2 k_0^2 
 \left\{|F(z-\bar w)|^2 - {1\over 2} (F(\bar z-w))^2 -
{1\over 2} (F(z-\bar w))^2 \right.\nn\\
&&\hspace{-1.0in}- F(w-\bar w)[F(z-w) - F(z-\bar w) - F(\bar z- \bar w)
+ F(\bar z- w)]\nn\\
&&\hspace{-1.0in}+\frac{1}{2}[F(z-w)F(z-\bar{w})+
F(\bar{z}-\bar{w})F(\bar{z}-w)\nn\\
&&\hspace{-1.0in}\left.-F(z-\bar{w})F(\bar{z}-\bar{w})-F(z-w)F(\bar{z}-w)]\right\}\nn\\
&&\hspace{-1.0in}\left.-\frac{1}{2}(\partial_{z}\partial_{w}P_{D}
+\partial_{\bar{z}}\partial_{\bar{w}}P_{D}
-\partial_{z}\partial_{\bar{w}}P_{D}-\partial_{\bar{z}}\partial_{w}P_{D})
\right]\,,\label{final}
\eeqa
where 
\beq
q\cdot h\cdot q = q^i h_{ij}q^j
\eeq
and
\beq
P_{D}=G(z,w)-G(z,\bar{w})\,,
\eeq
with $G(x_{1},x_{2})$ defined in eq. (\ref{G}). 

\section{Pinching limit}\label{pinch}
\renewcommand{\theequation}{6.\arabic{equation}}
\setcounter{equation}{0}

In the last section we obtained the final form of the amplitude, 
eq. (\ref{final}) for the scattering of a graviton and axion from two
parallel $0$-branes. We saw that the only contribution to this amplitude
comes from the odd spin structure. Our aim in this section is to make
contact with the field theory limit and to investigate the strongest
singularity for $q^{2}\rightarrow 0$. This will give us the leading
behaviour at large distances. We therefore analyze the behaviour of 
\beq
\int d^{2}z d^{2}w <V_{h}(k,z,\bar{z}) V_b(p,w,\bar{w})>\, 
\eeq
in the limit  $l\rightarrow \infty$ (field theory limit) and also 
$z\rightarrow w$ (pinching limit). 

We first analyze the limit $z\rightarrow w$. The
leading singularity in $q^2$ in this limit comes from the term 
$|F(z-w)|^2$. In 
Appendix C we show that this term away, from $w\sim z$, 
and also all the other terms  in the amplitude eq. 
(\ref{final}) give less singular results for $q^2\to 0$.

Clearly, upon using Eq. (\ref{exp}) and the definition of $F(z-w)$, eq.
(\ref{F}), and the behaviour of $\vartheta_{1}(z-w)$, one finds that the
contribution from the integration region $z\rightarrow w$ is of the form 
\beq
I=\int d^2z d^2w |F(z-w)|^2 e^{-k_\mu p_\nu <X^\mu(z) X^\nu(w)>} 
\stackrel{u=z-w\rightarrow 0}{\longrightarrow} \int d^2w \int 
d^2u\frac{|u|^{q^{2}}}{|u|^{2}}  
\left \vert {\vartheta'_1(0)\over \vartheta_1(w-\bar w)}
\right \vert ^{q^2} e^{{2\pi\over l}q^2(Imw)^2}\,.
\eeq
Converting to polar coordinates $u=\rho e^{i\theta}$ one can write this
integral as
\beq
\pi\int_{0}^{\Lambda^{2}}d\rho^{2}(\rho^{2})^{-1+\frac{q^{2}}{2}}
\left \vert {\vartheta'_1(0)\over \vartheta_1(w-\bar w)}
\right \vert ^{q^2} e^{{2\pi\over l}q^2(Imw)^2}\,,
\eeq
where the cutoff $\Lambda^{2}$ denotes a small integration region near
$z-w=0$. One can now do the $\rho^{2}$ integration and also the
integration over $Re w$, as the integrand is only a function of $Im w$, to 
obtain
 \beq
{2\pi\over q^2} (\Lambda^{q^2}) \int_0^l dIm w \left\vert {\vartheta'_1(0)
\over\vartheta_1(2i Im w)}\right \vert ^{q^2} e^{{2\pi\over l}q^2 (Im 
w)^2}\,. \eeq

Using the fact that 
\beq
\vartheta_1(z|2il) \stackrel{l\rightarrow\infty}{\longrightarrow} -2 
e^{{-\pi l\over 2}} sin(\pi z)
\eeq
one finds
\beq
\frac{\vartheta_{1}^{\prime}(0)}{\vartheta_{1}(2iImw)}\stackrel
{l\rightarrow\infty}{\longrightarrow}\frac{\pi}{sin(2\pi iImw)}\stackrel
{Imw\rightarrow\infty}{\longrightarrow}-2\pi ie^{-2\pi Imw}\,.
\eeq
The last limit follows since we are looking for the integration region 
where $Im w \sim l$.

Thus in this limit we find  
\beq
I={2\pi\over q^2} (2\pi\Lambda)^{q^2} \int_0^l d Im w e^{-2\pi q^2 Im w
(1-{Im w\over l})}\,.\label{I}
\eeq

Then substituting this behaviour in eq. (\ref{amp}), and keeping all the 
factors from eq. (\ref{final}), one finds that 
\beq
{\cal M}\stackrel{l\rightarrow \infty}{\longrightarrow}
\frac{a}{2}~q\cdot h \cdot q~q^2\int_{0}^{\infty}dl l^{-\frac{3}{2}}I\,.
\eeq

Now changing variables from $Imw$ to $\eta=(Im w)/l$ one finds that
\beqa
{\cal M}\rightarrow \pi ~a ~q\cdot h\cdot q 
\int_0^\infty
dl l^{-1/2} \int_0^1 d\eta e^{-2\pi l q^2 \eta (1-\eta)}\nn\\
= \pi~ a~ q\cdot h\cdot q~ {1\over \sqrt{q^2}}
\int_0^\infty dx x^{-1/2} \int_0^1 d\eta
e^{-2\pi x \eta (1-\eta)}\nn\\
= a~q\cdot h\cdot q~ \frac{1}{\sqrt{\vec{q}^2}}\cdot~ (constant),\label{ax}
\eeqa
noting that the integral over $\eta$ for large $x$ behaves as $1/x$,
and thus the subsequent integral is finite. We recall that $q^2$ is 
purely spacelike.

Since the limit $l\rightarrow \infty$ corresponds to the exchange 
of the lowest closed string states between the branes, the pinching 
limit graviton-axion amplitude has the correct momentum structure to be 
interpreted as the graph in figure 3, i.e the graviton and axion 
interact with the $0$-branes through the exchange of an intermediate axion 
which couples to the lowest states being exchanged between the branes. 
In fact, the pinching limit, as usual, selects the one particle exchange 
in the momentum transfer channel and the obvious candidate for this 
particle is the axion, whose coupling to the graviton through the energy 
momentum tensor corresponds to the right vertex in the diagram and to the 
structure $aq\cdot h\cdot q$. Thus in the  vertex at the left one sees the 
coupling of the axion with the lowest RR states exchanged between the 
branes. The propagators of these RR states carry only three momenta and 
no energy. The fact that the amplitude does not contain a pole like 
$1/\vec{q}^2$, but only $1/\sqrt{\vec{q}^2}$, suggests that the axion-RR-RR 
vertex is 
proportional to two powers of momenta. In fact, since the RR propagators 
behave as $1/\vec{k}^2$ and $1/(\vec{q}-\vec{k})^2$, integration over the 
$\vec{k}$, by dimensional reasons, will give $\sim\vec{q}^2/\sqrt{\vec{q}^2}$, 
which multiplied by the axion propagator $1/\vec{q}^2$ and the right vertex 
$aq\cdot h\cdot q$ reproduces our result for the amplitude.

 \medskip
\input epsf
\epsfxsize=10cm
\centerline{\epsffile{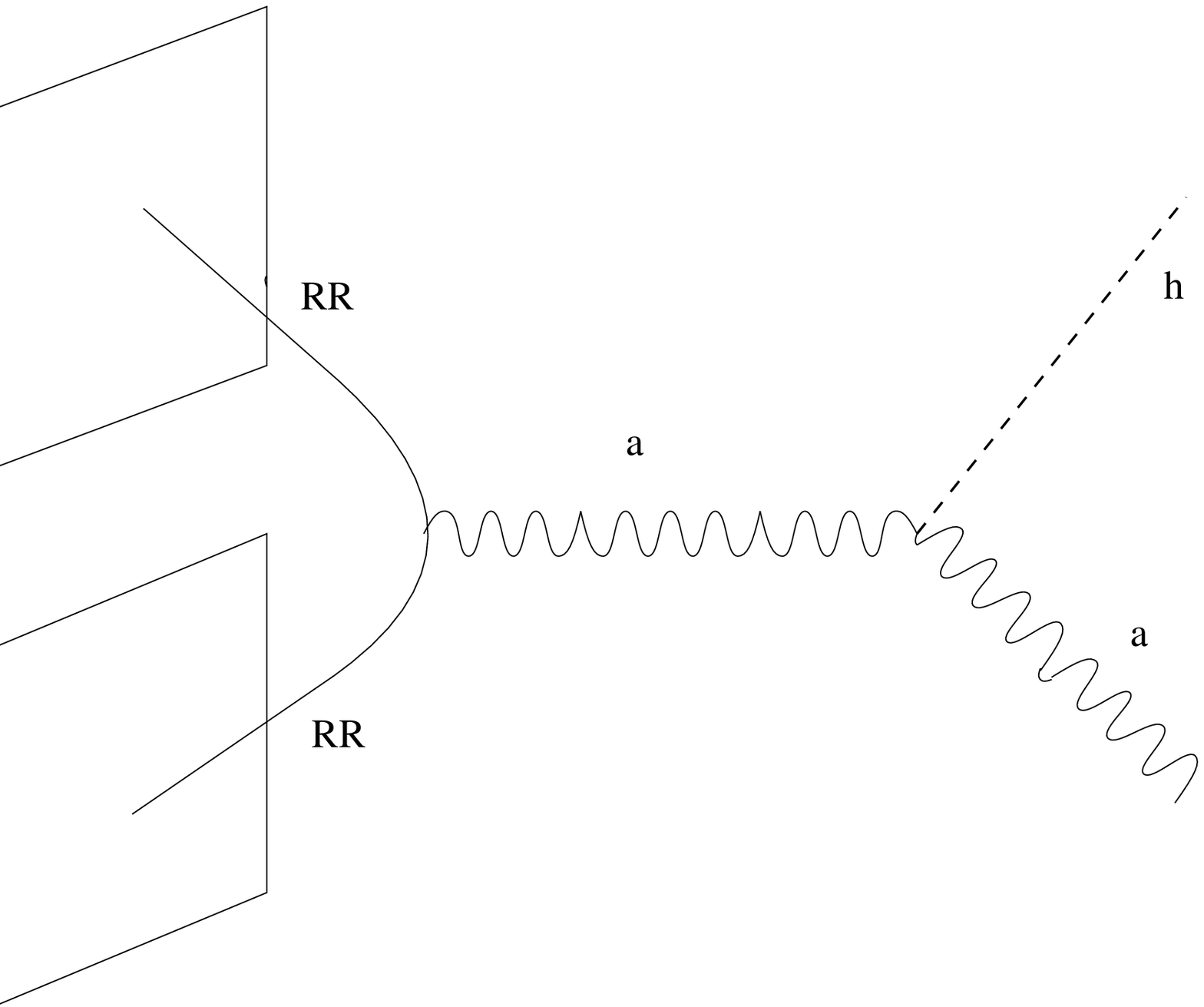}}
\centerline{Figure 3}

This interpretation of our result (let us stress that our result 
was obtained independently of the above field theory construction)  
may pose some problems. In fact it looks 
at variance with known rules from $N=2$ supersymmetric gauge theories, 
which are supposed to hold in the ``bulk", away from the branes, since 
the axion should be in a hypermultiplet and the RR states in a vector 
multiplet. It would seem that the well-known breaking of $N=2$ down to 
$N=1$ due to the branes sort of propagates away from them. This effect 
could be related to the fact that the RR states coming off the branes are 
necessarily off-shell since they carry zero energy, whereas one can verify 
that the four point amplitude $RR-h\rightarrow RR-a$ vanishes (and thus 
also its momentum transfer pole vanishes) for on-shell RR states in 
agreement with the above rules.

One notices also from eq. (\ref{ax}) that, if we make a three 
dimensional Fourier transform of the $1/\sqrt{\vec{q}^2}$ behaviour of the 
amplitude, we find a $1/r^2$ distribution for the static axion field at 
large distances from the source. This is to be contrasted with the 
normal behaviour of a scalar field which goes like $1/r$, and can 
be interpreted as due to the fact that the axion is coupled to a halo around
the pointlike sources rather than to the sources directly.

\vspace*{1cm}
{\bf Acknowledgements.}

We would like to thank E. Gava, T. Jayaraman, K.S. Narain,
A. Pasquinucci and H. Sarmadi 
for very useful discussions.

\vspace*{1cm}

\appendix{\bf Appendix A. D-brane vacuum amplitude}
\renewcommand{\theequation}{A.\arabic{equation}}
\setcounter{equation}{0}

In this Appendix we review the computation of the vacuum amplitude
for two parallel branes performed by Polchinski \cite{pol1} in order to
introduce notation.

The mode expansion of the bosonic field $X^\mu(\sigma_1,\sigma_2)$ is

\beqa
X^\mu(\sigma_1,\sigma_2) & = &x^\mu + i k^\mu \sigma_1 + {i \over 
{\sqrt{4\pi}}}  
\sum_n {1\over n} (\alpha_n^\mu e^{2\pi n(\sigma_1+i\sigma_2)}
+ \tilde\alpha_n^\mu e^{2\pi n(\sigma_1 - i\sigma_2)})\nn\\
&\equiv& x^\mu + i k^\mu \sigma_1 + X_{R}^{\mu}(\sigma_{1},\sigma_{2})
 +X_{L}^{\mu}(\sigma_{1},\sigma_{2})\,.\label{bosmod}
\eeqa
As is reviewed in Section~\ref{sec-amp}, the bosonic boundary 
conditions are
implemented, in the closed string formulation, by constructing a bosonic
boundary state $|B>_{B}$. Suppose to be general we have Neumann boundary
conditions for $X^{A}$ and Dirichlet for $X^{i}$. Then, if the vacuum is
defined as usual as $\alpha_n^\mu|0> = \tilde\alpha_n^\mu|0>=0$,
for $n>0$, it is seen that $|B>_{B}$  can be represented in terms of raising 
operators as

\beq
|B_{B}> = exp\{\sum_{n=1} ^\infty {1\over n} (-\eta_{AB} \alpha_{-n}^A 
\tilde\alpha_{-n}^B + \delta_{ij}\alpha_{-n}^i\tilde\alpha_{-n}^j)\} |0>\,.
\eeq

For the fermions, it is convenient to treat the NS-NS and R-R
sectors separately. Let us consider the NS-NS case first. 
The mode expansions are
\beq
\psi^\mu(\sigma_1,\sigma_2) = \sum_{m=-\infty}^{\infty} \psi^\mu_m 
e^{2\pi m(\sigma_1+i\sigma_2)},
\quad
\bar\psi^\mu(\sigma_1,\sigma_2) = 
\sum_{m=-\infty}^{\infty}\tilde \psi^\mu_m e^{2\pi m(\sigma_1-i\sigma_2)}\,,
\label{fermod}
\eeq
for $m$ half odd integer, and the boundary state
is fixed by the condition (using the conventions of ref. \cite{pol1})
\beq
(\psi_m^\mu 
+ i\eta  \tilde\psi^\mu_{-m} ) |B> = 0\,,\label{fbs}
\eeq

\noindent
where $\eta=\pm$ is 
 $+ (-)$ for Neumann (Dirichlet) boundary conditions.

If $\psi_m^\mu|0>=\tilde\psi_m^\mu|0>=0$ for half-odd $m>0$,
it is seen that the boundary state satisfying condition (\ref{fbs})
is

\beq
|B,\eta>_{NS} = exp\{- i \eta\sum_{m=1/2}^\infty \psi_{-m}\cdot
\tilde\psi_{-m}\}|0>
\eeq
 
 The GSO projection is performed by the operator
\beq
(-1)^F = -(-1)^{\sum_{m>0}\psi_{-m}\cdot\psi_m}
\eeq

\noindent
which acts on $|B,\eta>_{NS}$ as
\beqa
(-1)^F|B,\eta>_{NS} =  (-1)^{\tilde F}|B,\eta>_{NS}
= -|B,-\eta>_{NS}
\eeqa

Therefore, the GSO projected state is
\beq
|{\cal B}>_{NS} = {1\over 2} \{|B,\eta>_{NS} - |B,-\eta>_{NS}\}
\eeq

>From now on we consider the Light Cone (LC) expressions, 
thus keeping only the contributions of the coordinates
$\mu =2,...,9$.
Recalling the expression for the  Hamiltonian in terms of the oscillators 

\beqa
H &=& {k^2\over 2} +
 2\pi \left\{\sum_{n\geq 1} (\alpha_{-n}\cdot\alpha_n + \tilde\alpha_{-n}
\cdot\tilde\alpha_n)  + \sum_{m=1/2}^\infty m(\psi_{-m}\cdot\psi_m 
+ \tilde\psi_{-m}\cdot\tilde\psi_m)-1\right\}\nn\\
& = & {k^2\over 2}-2\pi + H_{B} +H_{F}\,,
\eeqa
one can compute the NS-NS contribution to the partition function ,
which, after integrating over $k$, can be expressed as
\beqa
Z_{NS}& = & 
\int_0^\infty dl 
 (2\pi l)^{-({{9-p}\over 2})} e^{-{Y^2\over 2l}}e^{2\pi l}
<B|e^{-lH_{B}}|B>_{B}\times\nn\\
&&\frac{1}{2}\{<B,\eta|e^{-lH_{F}}|B,\eta>_{NS}
-<B,\eta|e^{-lH_{F}}|B,-\eta>_{NS}\}\nn\\
& = & 8{\pi}^{4}V_{p+1}\int_0^\infty dl 
(2\pi l)^{-({{9-p}\over 2})} e^{-{Y^2\over 2l}}  
\vartheta'_1(0|2il)^{-4} \left [\vartheta_3(0|2il)^4 
- \vartheta_4(0|2il)^4\right ]\label{nspart}
\eeqa
where $V_{p+1} = <0|0>$ is the volume of the D-brane.

We recall the standard definitions
\beqa
\vartheta_2(z|2il)&=&\vartheta{1/2 \brack 0}(z|2il)\nn\\
\vartheta_3(z|2il)&=&\vartheta{0 \brack 0}(z|2il)\nn\\
\vartheta_{4}(z|2il)&=&\vartheta{0 \brack 1/2}(z|2il)\,.
\eeqa
In a similar way we can proceed in the R-R sector.
In this case, the mode expansions are the same as in eq.(\ref{fermod}) 
but with $m$ integer.
Here,  the vacuum is
degenerate 
due to the zero modes 
$\psi_0^\mu, \tilde\psi_0^\mu$. 

The boundary state satisfying
\beq
(\psi^\mu_m +i \eta \tilde\psi^\mu_{-m})|B,\eta>_R=0
\eeq
for every $m$ including $m=0$ is 
\beq
|B,\eta>_R = exp\left\{-i\eta\sum_{m\ge 1}\psi_{-m}\cdot \tilde\psi_{-m}
\right\}
|0,\eta>
\eeq

 In this sector, the
GSO projection is performed by the operator
\beq
(-1)^F = \psi_0^{11} (-)^{\sum_{m>0}\psi_{-m}\cdot\psi_m}
\eeq
which yields
\beq
|{\cal B}>_R = {1\over 2} (|B,\eta>_R - |B,-\eta>_R)
\eeq

It is seen that the term corresponding to the odd spin structure is
$<B\eta |e^{-H_{F}l}|B-\eta >_R=0$ (it is proportional to
$\vartheta_1 (0)^4=0$). Here $H_{F}$ is the Hamiltonian for the RR fermions.
Repeating the calculation above (i.e. Eq.(\ref{nspart}) 
with the appropriate Hamiltonian)
it is seen that

\beq
Z_{R} = 8{\pi}^{4}V_{p+1}\int_0^\infty 
(2\pi l)^{-({{9-p}\over 2})} e^{-{Y^2\over 2l}}
\vartheta'_1(0|2il)^{-4}\vartheta_2(0|2il)^4
\eeq

In order to make the algebraic sum of the NS-NS and RR sectors, we 
take the same signs which hold for the partition function on the torus. 
Thus, 
the full partition function is:
 \beq Z= 8{\pi}^{4}V_{p+1}\int_0^\infty 
(2\pi l)^{-({{9-p}\over 2})} e^{-{Y^2\over 2l}}
\vartheta'_1(0|2il)^{-4} \left [\vartheta_2(0|2il)^4 - \vartheta_3(0|2il)^4
+ \vartheta_4(0|2il)^4 \right]\,.\label{pf}
\eeq
It vanishes due to the abstruse identity, reflecting the fact that there is
no net force between BPS states. 

\appendix{\bf Appendix B.  D-brane vacuum amplitude  on  orbifolds}
\renewcommand{\theequation}{B.\arabic{equation}}
\setcounter{equation}{0}

In this Appendix we consider an orbifold compactification to four dimensions.
The standard $Z_3$ orbifold in string theory
 is known to break some of the supersymmetries due to the inclusion of
twisted boundary conditions.
Thus it is interesting to see whether the force between BPS states
is affected by this more realistic setup.

Now the compactified fields are propagating along the cylinder. It is possible
to construct the boundary states in this case, taking into account the
twisted sectors  provided by the orbifold construction.  
Consider the discrete group $G=Z_3$ that acts on the
coordinates $X^\mu$, $\psi^\mu$ as $\pm2\pi /3$ rotations on pairs
of them.
An orbifold is constructed by compactifying the coordinates
$\mu =4,5,6,7,8,9$ on a 6-torus and 
by identifying points that are
equivalent up to an element $g$ of $G$, i.e.
\beq
X^\mu(\sigma_1,\sigma_2 = 1) = g X^\mu(\sigma_1,\sigma_2 = 0)
\eeq
This is possible if the mode $x^{\mu}$ is a fixed point, 
i.e. $x^\mu=g x^\mu$. For $g\not= 1$ we call this case a twisted sector
in $\sigma_2$.

This divides the Hilbert space into sectors where the only change is in the
boundary conditions. Moreover,
since $G$ is a symmetry of the Hamiltonian, we must project onto $G$-invariant
states with the projector $\sum_k g^k$, both for the twisted and 
untwisted sectors in $\sigma_2$. This projector acts similarly to 
the GSO projection, that is given a boundary state $|B>$ we 
have to form $\sum_k g^k|B>$. 

In order to construct the boundary states, let us consider Dirichlet
boundary conditions on the bosonic fields, i.e. $X^\mu(\sigma_1=0,\sigma_2)=
Y_1^\mu, X^\mu(\sigma_1=l,\sigma_2) = Y_2^\mu$, where $Y_1^\mu$ and
$Y_2^\mu$ are the positions of the branes. (If the twisted sector in $\sigma_2$
is considered, then $Y_1$ and $Y_2$ must be fixed points of the orbifold).
These conditions can be written (in the notation of eq. (\ref{bosmod}.
Appendix A) as
\beq
\partial_{\sigma_2} X_R\vert_{\sigma_1=0} = - \partial_{\sigma_2}
X_L\vert_{\sigma_1 = 0} ~~~~~~~~~
\partial_{\sigma_2} X_R\vert_{\sigma_1=l} = - \partial_{\sigma_2}
X_L\vert_{\sigma_1 = l}\label{eq2}
\eeq
Similarly for Neumann boundary conditions we have
\beq 
\partial_{\sigma_1} X_R\vert_{\sigma_1=0} = - \partial_{\sigma_1}
X_L\vert_{\sigma_1 = 0}  ~~~~~~~~~
\partial_{\sigma_1} X_R\vert_{\sigma_1=l} = - \partial_{\sigma_1}
X_L\vert_{\sigma_1 = l}\label{eq3}
\eeq
In this way we can construct the states $|B>$ which only contain the
nonzero modes. Now, for every state we have to project over the $Z_3$ invariant
content. This is performed by applying $\sum_{m,n} g^m \tilde g^n$, where
$g$ and $\tilde g$ are elements of $Z_3$ and act on the left and right movers
respectively. This implies that the condition to be satisfied by the 
$Z_3$ projected boundary
states will be deformed as
\beqa
g^m \partial_{\sigma_2} X_L\vert_{\sigma_1=0} = - \tilde{g}^n \partial_{\sigma_2}
X_R\vert_{\sigma_1 = 0} ~~~~~~~~~
g^{m'}\partial_{\sigma_2} X_L\vert_{\sigma_1=l} = - \tilde{g}^{n'} \partial_{\sigma_2}
X_R\vert_{\sigma_1 = l}\nn\\
g^m \partial_{\sigma_1} X_L\vert_{\sigma_1=0} = - \tilde{g}^n \partial_{\sigma_1}
X_R\vert_{\sigma_1 = 0} ~~~~~~~~~
g^{m'}\partial_{\sigma_1} X_L\vert_{\sigma_1=l} = - \tilde{g}^{n'} \partial_{\sigma_1}
X_R\vert_{\sigma_1 = l}
\eeqa
and one has to sum over all the possibilities. Since the partition function
depends only on the relative phase of $g^m \tilde{g}^n$ with respect to 
$g^{m'} \tilde{g}^{n'}$
and further of $g^{m'}$ with respect to $\tilde{g}^{n'}$, the Dirichlet and Neumann
boundary conditions on the projected states can be summarized respectively as
\beqa
\partial_{\sigma_2} X_R\vert_{\sigma_1=l} = - g \partial_{\sigma_2}
X_L\vert_{\sigma_1 = l}\nn\\
\partial_{\sigma_1} X_R\vert_{\sigma_1=l} = - g \partial_{\sigma_1}
X_L\vert_{\sigma_1 = l}\label{tbc}
\eeqa
for the states   $|Bg>\equiv g|B>$ on the brane at 
$\sigma_1 =l$. 
Thus, the operation $g$ induces a twist in the 
$\sigma_1$ direction.

Notice that, alternatively, one could interprete eq.(\ref{tbc})
in the following way: due to the identifications introduced by
the orbifold, it is also possible to implement the boundary conditions
eqs.(\ref{eq2}) or (\ref{eq3}) 
by equating the derivative of the right movers to (minus) 
the derivative of the twisted left movers.

>From now on we will consider Dirichlet boundary conditions in 
every space coordinate. Thus we are in the case of the D$0$-brane. 
Actually, we would get the same Light Cone
partition function also with Neumann conditions in the compactified 
directions - 
 taking the orbifold fixed points would seem anyhow to 
imply the Dirichlet ones; however, in the orbifold case, 
the integration over the compactified positions, which would 
be done in the Neumann case, means also a discrete sum over the
fixed points.   
Thus when we say $0$-brane, we make essentially reference to the
 uncompactified coordinates.

Working in light-cone coordinates $\mu,\nu = 2,...,9$,
let us consider the compactified
directions $4,...,9$.
It is convenient to introduce complex fields
\beq
X^{4,5} = X^4 + i X^5 \qquad \bar X^{4,5} = X^4 - i X^5
\eeq
(and similarly $X^{6,7}$, $X^{8,9}$). 
The eigenvalues of $g$ acting on these complex fields are
 
\beq
g = exp \{2\pi i (z_{4} + z_{6} + z_{8})\}
\eeq
where $z_{4} \pm  z_{6} \pm z_{8} = 0$ mod $2$.
(see \cite{minahan}). In order to preserve one supersymmetry, $G$ is assumed
to be an abelian subgroup of $SU(3)$. Therefore, for a
$Z_3$ orbifold,  $z_{a}=n_{a}/3$ and
$n_{4}\pm n_{6}\pm n_{8}=6n$.

We first discuss the untwisted sector in $\sigma_2$.

Let us consider $X^{4,5}$.
The corresponding oscillator modes
\beq
\beta_n = \alpha_n^4 + i \alpha_n^5, \quad
\beta_n^* = \alpha_n^4 - i \alpha_n^5
\eeq
with
\beq
[\beta_n, \beta_l] = 0, \quad [\beta_n, \beta^*_{-l}] = 2 n \delta_{lm}
\label{bc}
\eeq
and $\tilde{\beta}_{n},\tilde{\beta}^{*}_{n}$ defined in a similar way for
the left movers, 
yield the following condition on the boundary state 
\beq
(\beta_n + g_{4} \eta \tilde\beta_{-n})|B_{4},g_{4},\eta> = 0\,.
\label{betcon}
\eeq
Notice that we are using $\eta$ to distinguish Neumann and Dirichlet
boundary conditions. This is meant to stress the similarity between
the projection over $Z_3$ and GSO invariant states. As far as we consider
branes with the same Neumann and Dirichlet coordinates, 
the bosonic content of the bra and ket boundary states
on both branes will coincide and $\eta$ will only refer to the GSO projection.

It is thus seen that
\beq
|B_{4},g_{4},\eta> = exp\{-\eta\sum_{n \ge 1}{1\over 2n} (g^*_{4}
\beta_{-n}  
\tilde\beta^*_{-n} + g_{4} \beta^*_{-n}\tilde\beta_{-n})\}|0>
\eeq

It is now possible to compute the bosonic contribution to the vacuum
amplitude, taking into account the expression for the Hamiltonian in
terms of the new oscillators, namely
\beq
H= \pi \sum_n (\beta_{-n}\beta^*_{n} + \beta^*_{-n} \beta_n + \tilde\beta_{-n}
\tilde\beta^*_n + \tilde\beta^*_{-n}\tilde\beta_n)
\eeq

One then gets
\beq
<B_{4},g_{4},\eta|e^{-lH}|B_{4},g'_{4},\eta> =
\prod_{n=1} \left \vert {1\over 1-g^*_{4}g'_{4}e^{-4 \pi l n}}\right 
\vert ^2 =
\prod_{n=1}\left [ 1 + q^{4n} - 2 {\rm cos} (2 \pi z_{4}) q^{2n}\right ] ^{-1}
\eeq
where we have introduced $q=e^{-2\pi l}$ and $g_{4}^*g'_{4} = 
e^{2\pi i z_{4}}$.

In terms of Jacobi theta functions this expression can be rewritten as
\beq
<B_{4},g_{4},\eta|e^{-lH}|B_{4},g'_{4},\eta> =
{{2 f(q^2) q^{1/4} sin (\pi z_{4})}\over {\vartheta_1(z_{4}|2il)}}\label{zb4}
\eeq
where
\beq
f(q^2) = \prod_{n=1}(1-q^{2n}) = \left ({\vartheta_1'(0|2il)\over {2\pi q^{1/4}}}
\right )^{1/3}
\eeq

Taking now into account
all the contributions from the compactified directions as well as
the (2,3) spacetime sector and the normal ordering term in the hamiltonian
($q^{-2/3}$),
it is seen that the bosonic sector produces (in the notation of 
eq. (\ref{partfunct})) 
\beq
Z^B_{LC} = 
\left [ 2 f(q^2) \right ]^4{\pi q^{1/3} \over {\vartheta'_1(0|2il)}} 
\prod_{a} {{sin (\pi z_{a})}\over {\vartheta_1({z_{a}}|2il)}}\,.
\eeq

Let us now consider the NS fermionic sector.  
Complex fermions are defined as
\beq
\chi^{(4,5)} = \psi^4 + i\psi^5, \quad \chi^{(4,5)*} = \psi^4 - i \psi^5
\eeq
(and similarly for $\chi^{(6,7)}$ and $\chi^{(8,9)}$ and the left movers
$\tilde{\chi},\tilde{\chi}^{*}$) with
\beqa
\{\chi_n,\chi_m\} = \{\chi_n,\tilde\chi_m\} = \{\tilde\chi_n,\tilde\chi_m\}
= 0 ~~~~
\{\chi_n, \chi^*_{-m}\} = \{\tilde\chi_n, \tilde\chi^*_{-m}\} = 2 \delta_{nm}\,.
\eeqa

The condition to be satisfied by the twisted boundary state is
\beq
(\chi_n + i g  \eta \tilde\chi_{-n})|B,g,\eta> = 0\label{fcond}
\eeq
(where we avoid repeating the pairs of indices $(4,5),(6,7),(8,9)$ on the
fields, the twists and the states). 
It is seen that the state satisfying condition (\ref{fcond})
is
\beq
|B,g,\eta> = 
exp\left \{-{\eta i\over 2}\sum_n (g\chi^*_{-n}\tilde\chi_{-n}
+ g^* \chi_{-n} \tilde\chi^*_{-n})\right \}|0>\,.
\eeq
Taking into account all the compactified directions as well as the 
(2,3) spacetime
sector, the full LC boundary state is
\beq
|{\cal B}> = |B,\eta>_{23} |B, g_{4},\eta_{4}>|B, g_{6},\eta_{6}>
|B, g_{8},\eta_{8}>\,.
\eeq

Now the vacuum amplitude can be computed, using the expression for
the Hamiltonian in terms of the complex fields,
\beq
H=\pi\sum_{m=1/2}^\infty m(\chi_{-m}\chi^*_m + \chi_{-m}^*\chi_m + 
\tilde\chi_{-m}\tilde\chi^*_m + \tilde\chi^*_{-m}\tilde\chi_m)\,.
\eeq
Proceeding similarly as in the bosonic case, we find
 (with $g^*g'=e^{2\pi iz_{a}}$) 

\beqa
<B,g,\eta|e^{-l H}| B,g',\eta '> = ~~~~~~~~~~~~~~~
~~~~~~~~~~~~\nn\\
~~~~~~~~= \prod_n\left |1 +  \eta \eta'
g^*g' e^{-4\pi l n}\right |
= \prod_{n=1}^\infty \left |1\pm e^{2\pi iz_{a}} q^{(2n-1)}\right |^2
\eeqa
(Note that $\eta \eta' = \pm$ for the two possible cases 
of the GSO projection). In order to compute the partition function
we have to
put together
all the compactified directions as well as the spacetime contribution,
perform the GSO projection and include the normal ordering term
in the hamiltonian ($q^{-1/3}$). Finally the contribution  in the NS sector
amounts to, in the notation of eq. (\ref{partfunct}),
\beq
Z^F_{LC(NS\eta\eta '=1)}-Z^F_{LC(NS\eta\eta '=-1)}
 =
{{\vartheta_3(0|2il)\prod_{a}\vartheta_3(z_{a}|2il)}\over 
q^{1/3}[f(q^2)]^4}
-{{\vartheta_4(0|2il)\prod_{a}\vartheta_4(z_{a}|2il)}\over 
q^{1/3}[f(q^2)]^4}\,. \eeq

Let us now turn our attention to the R sector. In this case the $n=0$
modes deserve a special treatment, which we have discussed in Section
\ref{sec-boundary}.
We recall from that Section that the $n=0$ contribution in the
odd spin structure case gives a vanishing result after the projection
 $\sum_k g^k|B>$ is taken into account. Instead, the $n=0$ mode
part for the even spin structure case does not vanish. Rather,
it gives a factor $\prod_{a}2cos(\pi k z_{a})$ for each 
$g^k,k\not= 0$.

The $n\not= 0$ contribution in this sector, for a pair of 
compactified coordinates,  amounts to
\beqa
<B|e^{-lH}|g^kB>_{\eta \eta' = 1}  = 
\prod_{n=1}^\infty \left \vert 1 + e^{2\pi iz_{a}} q^{2n}\right \vert ^2\nn\\
~~~~~~~~~~~~~~~~~~~~~~~~~~~~ = \prod_{n=1}^\infty [1 + q^{4n} + 2 q^{2n} 
{\rm cos}(2\pi z_{a})]
\eeqa
which in terms of Jacobi theta functions is
\beq
{\vartheta_2(z_{a}|2il)\over {2 f(q^2) q^{1/4} {\rm cos}(\pi z_{a})}}
\eeq

Putting together the  contributions from the (2,3) spacetime fermions
and the internal $(4,5), (6,7), (8,9)$ directions, after including
the zero point energy factor $q^{2/3}$ we get
\beqa
Z^F_{LC(R\eta\eta '=1)}=
 {\vartheta_2(0|2il)\prod_{a}\vartheta_2(z_{a}|2il)\over {q^{1/3} 
f(q^2)^4}}\,.   \eeqa

In order to make the algebraic sum of 
 the R and NS sectors, we take the same signs which hold for 
the partition function on the torus. Thus we get the $LC$
partition function
\beqa
\lefteqn{Z_{LC}=Z_{LC}^B\cdot 
(Z^F_{LC(R\eta\eta '=1)}-Z^F_{LC(NS\eta\eta '=1)}+
Z^F_{LC(NS\eta\eta '=-1)})=} 
\nn\\ 
& &{16 \pi\over \vartheta_1'(0|2il)} \prod_{a}{sin(\pi z_{a})\over 
\vartheta_1(z_{a}|2il)}\nn\\
& &\times \left \{ \vartheta_2(0|2il)\prod_{a}\vartheta_2(z_{a}|2il) -
 \vartheta_3(0|2il)\prod_{a}
\vartheta_3(z_{a}|2il) + \vartheta_4(0|2il)\prod_{a}\vartheta_4
(z_{a}|2il)\right \}\,.
\eeqa
which vanishes due to the Riemann identity.

When the position of the brane is
 on the fixed point of the orbifold one has also to
include the twisted sector in $\sigma_2$.
In this case, the fields in the compactified
dimensions may be diagonalized such that
\beq
X^{a,b}(\sigma_2 + 1) = e^{2\pi i z_{a}} X^{a,b}(\sigma_2),\quad  X^{*a,b}(\sigma_2+1) =
e^{-2\pi i z_{a}} X^{*a,b}(\sigma_2)
\eeq
Therefore, the mode expansion of $X^{a,b}$ and $X^{*a,b}$ is given by
\beqa
X^{a,b} = x^{ab} + \frac{i}{\sqrt{4\pi}}\sum_{n\in Z}\left [ \frac{1}{\sqrt{n-1/3}} \gamma_n e^{-2\pi i (n-1/3)\sigma_2}
+ \frac{1}{\sqrt{n-2/3}} \tilde \gamma_n e^{2\pi i(n-2/3)\sigma_2}\right ]\nn\\
X^{*a,b} = x^{*ab} +\frac{i}{\sqrt{4\pi}}\sum_{n\in Z}\left [ \frac{1}{\sqrt{n-2/3}} \gamma^*_n 
e^{-2\pi i (n-2/3)\sigma_2}
+ \frac{1}{\sqrt{n-1/3}} \tilde \gamma^*_n e^{2\pi i(n-1/3)\sigma_2}\right ]\nn
\eeqa
where
\beqa
\gamma_n|0> = 0, \quad \tilde\gamma_n|0> = 0, \quad
\gamma^*_n|0> = 0, \quad \tilde\gamma^*_n|0> = 0
\eeqa
for $n>0$. Therefore $\gamma_n^{\dag} = \gamma^*_{-n+1}$ and
$\tilde\gamma^{\dag}_n = \tilde\gamma^*_{-n+1}$ and the commutation
relations are
\beq
[ \gamma_n, \gamma^*_{-l+1} ] = 2 \delta_{nl},\quad  n,l >  0, \quad
[ \gamma_{l+1} , \gamma^*_n ] = 2 \delta_{nl},\quad  n,l\leq 0\,. 
\label{nc}
\eeq

In terms of the new oscillators, the Hamiltonian is
\beq
H = \pi\sum_{n=0}\left [ (n+{1\over 3})(\gamma_{-n}\gamma^*_{n+1}
+ \tilde\gamma^*_{-n}\tilde\gamma_{n+1}) + (n+{2\over 3}) (\gamma^*_{-n}
\gamma_{n+1} + \tilde\gamma_{-n}\tilde\gamma^*_{n+1})\right ]
\eeq
and the condition to be verified by the boundary state is
\beq
(\gamma_n + \eta g \tilde\gamma_{-n+1})|B> = 0\,.
\eeq

It is easy to see that the boundary state  verifying it is
\beq
|B> = exp\{-{\eta\over 2}\sum_{n=1}(g \gamma^*_{-n+1}\tilde\gamma_{-n+1}
+ g^*\gamma_{-n+1}\tilde\gamma^*_{-n+1})\}|0>
\eeq

Therefore, the contribution from this sector to the partition function
is
\beqa
\lefteqn{<B,g,\eta|e^{-lH}|B,g',\eta> =} \nn\\ 
& &\prod_{n=1}\left ( 1 - g^* g' e^{-4\pi l(n-{1\over 3})}\right )^{-1}
\left ( 1 - g g'^* e^{-4\pi l (n-{2\over 3})}\right )^{-1}
\eeqa
 with ($g^*g'=e^{2\pi iz_a}$).

After putting all the contributions from the spacetime and internal
directions together, the full bosonic part of the path integral is
in the notation of eq.(\ref{partfunct})
\beq
Z^B_{LCh} =
\prod_{a}\prod_{n=1}\left [1-e^{2\pi iz_{a}} e^{-4\pi l(n-{1\over 
3})}\right ]^{-1}
 \left [1-e^{-2\pi iz_{a}} e^{-4\pi l(n-{2\over 3})}\right ]^{-1}\,,
\eeq
where the subscript $h$ means twist in the $\sigma_2$ direction.

Using the infinite product expansion of the generalized 
$\vartheta$-functions, \beqa
\vartheta{a\brack b}(z|2il) = q^{a^2} e^{2\pi i a(b+z)} \prod_n (1 - q^{2n})
\prod_n(1 + q^{2n-1} e^{2\pi i (2ial + b + z)})\nn\\
\prod_n(1 + q^{2n-1} e^{-2\pi i (2ial + b + z)})\label{gtf}
\eeqa
$Z_{LCh}^B$ can be expressed as
\beq
Z_{LCh}^B = e^{\pi i/2} f(q^2) q^{1/12} \left [
\prod_{a}\vartheta{1/6 \brack 1/2}(z_{a}|2il)\right ]^{-1} 
\eeq

In order to discuss the fermionic contribution, let us start with
the NS sector. The condition (\ref{fcond}) is now modified to
\beq
(\chi_n + i g \eta \tilde\chi_{-n+1})|B,g,\eta> = 0
\eeq
which is satisfied by
\beq
|B,g,\eta> = exp\{-{i\eta\over 2}\sum_{n=1}(g\chi^*_{-n+1}
\tilde\chi_{-n+1} + g^*\chi_{-n+1}\tilde\chi^*_{-n+1})\}|0>
\eeq

The corresponding Hamiltonian is
\beq
H = \pi\sum_{n=0}\left [ (n+{5\over 6})(\chi_{-n}\chi^*_{-n+1}
+\tilde\chi^*_{-n}\tilde\chi_{-n+1}) + (n+{1\over 6})(\chi^*_{-n}
\chi_{-n+1} + \tilde\chi_{-n}\tilde\chi_{-n+1})\right ]
\eeq
and therefore, each complex pair of directions contributes
\beq
Z^F_{LCh(NS\eta\eta ')}\equiv
<B,g,\eta|e^{-lH}|B,g',\eta '> = \prod_{n=1}\left [ 1 +
\eta \eta' g^*g' e^{-4\pi l(n-{5\over 6})}\right ]
\left [ 1 + \eta \eta' g g'^* e^{-4\pi l (n-{1\over 6})}\right ]
\eeq

Taking into account the spacetime and the three pairs
of complex directions, and using Eq.(\ref{gtf}), 
the full contribution of the NS sector can be written as
\beqa 
\lefteqn{Z^F_{LCh(NS\eta\eta'= 1)} + Z^F_{LCh(NS\eta\eta'= -1)} =}\nn\\
& & q^{-1/3} f(q^2)^{-4} 
\left\{\vartheta{0\brack 0}(0|2il) \prod_{a}\vartheta{-1/3 \brack 0} 
(z_{a}|2il) - \vartheta{0 \brack 1/2 }(0) \prod_{a}\vartheta{-1/3 
\brack 1/2} (z_{a}|2il)\right\}\nn\\
& & = f(q^2)^{-4} 
\left\{\vartheta{0\brack 0}(0|2il) \prod_{a}\vartheta{0 \brack 0} 
(z_{a}-2il/3|2il) + \vartheta{0 \brack 1/2 }(0) 
\prod_{a}\vartheta{0 \brack 1/2} (z_{a}-2il/3|2il)\right\}\,.\label{zns}
\eeqa 
Recall that in the twisted sector for the $Z_3$ orbifold there has to be a 
relative positive sign between the $\eta\eta'=1$ and the 
$\eta\eta'=-1$ sectors because of invariance under $\tau \rightarrow\tau + 3$. 
 
Let us now consider the R sector. 
In the $\sigma_2$ twisted sector there are no $n=0$ modes 
in the compactified dimensions. In order to deal with the
$n=0$ modes of the uncompactified coordinates we put 
for them a small twist in the $\sigma_1$ direction  
, i.e. $g^*g' = e^{2\pi i \epsilon}$,
otherwise the odd spin structure case $\eta\eta '=-1$ would
be identically zero. 
Thus we write for the (2,3) spacetime directions 
\beq
<B,\eta|e^{-l H_0}|B,\eta'>_{(2,3)} =
\left ( 1+\eta\eta' e^{2\pi i \epsilon}\right ) \prod_{n=1} \left
\vert 1+\eta\eta'  e^{-4\pi l n}\right \vert^2
\eeq
In  the case of $\eta \eta'=1$ (even spin structure) we can
directly take the limit $\epsilon\to 0$

\beq
<B,\eta|e^{-l H_0}|B,\eta>_{(2,3)}= 
{\vartheta_2(0)\over q^{1/4} f(q^2)}   
\eeq
and thus in the notation of eq. (\ref{partfunct})

\beqa
Z^F_{LCh(R\eta\eta '=1)} 
&=&{\vartheta_{2}(0)
\over q^{1/4}f(q^2)} \prod_{a}\prod_{n=1}
(1 + e^{2\pi iz_{a}} e^{-4\pi l (n-{1\over 3})}) 
 (1 +  e^{-2\pi iz_{a}}e^{-4\pi l (n-{2\over 3})})  \nn\\
&=& {\vartheta{ 1/2 \brack 0}(0|2il) \over {q^{1/3} f(q^2)^4}}
\prod_{a}\vartheta{1/6 \brack 0}(z_{a}|2il)\nn\\
&=& {\vartheta{ 1/2 \brack 0}(0|2il) \over {f(q^2)^4}}
\prod_{a}\vartheta{1/2 \brack 0}(z_{a}-2il/3|2il)\,.\label{zr}
 \eeqa

In the odd spin structure case $\eta \eta'=-1$
we have for small $\epsilon$
\beq
<B,\eta|e^{-l H_0}|B,\eta '>_{(2,3)} 
\rightarrow {i\epsilon \vartheta^{'}_1(0)\over q^{1/4}
f(q^2)} \eeq
In this case 
the $n\not= 0$ modes contribute the inverse of the bosonic sector
(compare eq.(\ref{zb4})  for $z_{4}\to\epsilon$), and
thus we have in the notation of eq. (\ref{partfunct}) 
\beqa
Z^B_{LC}\cdot Z^F_{LC(R\eta\eta '=-1)}=
-2\pi i\epsilon \sim
 -2\pi i 
{\vartheta_1(\epsilon )\over \vartheta'_1(0)}\,,\label{odd}
\eeqa
which is zero for $\epsilon\to 0$. But of course, as we have seen in
Section \ref{sec-amp}, the insertion of vertices in the amplitude can give a 
nonzero
result also for the odd spin structure, and we are indeed interested in
this case. To be precise, one should go beyond the light cone and
include also the $(0,1)$ and $(\beta\gamma )$ ghost contribution,
with their zero modes. Altogether this gives the rules  at 
the end of Section \ref{sec-amp} (formally, the insertion of $\psi^2\psi^3$
,for instance, would provide a factor proportional to
${ \vartheta'_1(0)\over \vartheta_1(\epsilon )}$ to multiply eq. 
(\ref{odd})).

We conclude the computation of the  $LC$ partition function by 
assembling its even spin structure part.
We take the standard combination, see eq.(43),
\beqa
Z_{LCh}=Z_{LCh}^B\cdot
(Z^F_{LCh(NS\eta\eta '=1)}+Z^F_{LCh(NS\eta\eta '=-1)} -
Z^F_{LCh(R\eta\eta '=1)})\label{ztwist}
\eeqa
We can now show that this vanishes for each twisted sector. We see from 
eqs. (\ref{zns}) and (\ref{zr}) that the above equation is proportional 
to the sum 
\beqa
\vartheta{0\brack 0}(0|2il) \prod_{a}\vartheta{0 \brack 0} 
(z_{a}-2il/3|2il) + \vartheta{0 \brack 1/2 }(0) 
\prod_{a}\vartheta{0 \brack 1/2} (z_{a}-2il/3|2il)\nn\\
- \vartheta{ 1/2 \brack 0}(0|2il)
\prod_{z_{a}}\vartheta{1/2 \brack 0}(z_{a}-2il/3|2il)\,.
\eeqa

Writing $w_1=z_{4}-2il/3\,\,,w_2=z_{6}-2il/3\,\,,w_3=z_{8}+4il/3$
one can convert the above expression to
\beqa
e^{2\pi i(z_{8}+il/3)}\left\{\vartheta{0\brack 0}(0|2il) 
\prod_{a}\vartheta{0 \brack 0} (w_{a}|2il) - \vartheta{0 \brack 1/2 }(0) 
\prod_{a}\vartheta{0 \brack 1/2} (w_{a}|2il)\right.\nn\\
\left.- \vartheta{ 1/2 \brack 0}(0|2il)
\prod_{a}\vartheta{1/2 \brack 0}(w_{a}|2il)\right\}\,,
\eeqa
which vanishes due to the Riemann identity as $\sum_{a}w_a = 0$ mod $2$.

Finally, let us make a comment on the boundary conditions. In the
case of Neumann boundary conditions in the compactified directions,
we have to integrate over the branes' position and this includes a sum
over the orbifold fixed points. Thus, it is like taking Dirichlet 
conditions on those fixed points. In the case we start by considering
Dirichlet conditions in the 
compactified directions, we should take the branes at
rest at a generic fixed position 
and work out the dynamics of the closed string. But we have further
to consider that in the low energy state of the brane the wave
function will be spread over the compactified directions. Thus we have
to do an average over the compactified position of the brane,
as it would be done for the nuclei wave function in a molecule
within the Born Oppenheimer approximation. Like above,  
the integration
over the position includes also a sum over the fixed points. In conclusion,
the fixed points always contribute, bringing in 
the $\sigma_2$-twisted sector.

\appendix{\bf Appendix C. Even spin structures contribution to the 
axion-graviton scattering}

\renewcommand{\theequation}{C.\arabic{equation}}
\setcounter{equation}{0}

In this appendix we discuss the contribution from the even spin 
structures to the graviton-axion scattering amplitude.
The only possible non vanishing invariant in this case 
is $(p\cdot h\cdot b \cdot k)$. This can arise either from purely 
bosonic contractions or from terms containing four or eight fermions. 
Actually the contribution from the eight fermionic contractions 
is found to vanish.  The result is
\beqa
(p\cdot h\cdot b \cdot k)\times e^{-<k\cdot X(z) p\cdot X(w)>}
 \{\partial_z\partial_wP_D \bar\partial_zP_D
\bar\partial_wP_D - \bar\partial_z\bar\partial_wP_D \partial_zP_D
\partial_wP_D - \partial_z\bar\partial_wP_D \bar\partial_zP_D
\partial_wP_D\nn\\
 + \bar\partial_z\partial_wP_D \partial_zP_D
\bar\partial_wP_D + \partial_z\partial_wP_D F(\bar z - \bar w)^2_s -
\bar\partial_z\bar\partial_wP_D F( z -  w)^2_s \nn\\
- \partial_z\bar\partial_wP_D F(\bar z -  w)^2_s +
\bar\partial_z\partial_wP_D F( z - \bar w)^2_s +
\partial_zP_D \partial_wP_D (k\cdot p) F(\bar z - \bar w)^2_s \nn\\
-\bar\partial_zP_D \bar\partial_wP_D (k\cdot p) F( z - w)^2_s -
\partial_zP_D \bar\partial_wP_D (\tilde k\cdot p) F(\bar z -  w)^2_s
+ \bar\partial_zP_D \partial_wP_D (\tilde k\cdot p) F( z - \bar w)^2_s \}
\eeqa
Here $F(x_1-x_2)_s$ are any of the three even spin structure fermionic 
propagators. As we see, the terms come out in pairs,
always in the form: a term minus its complex conjugate.
Now, it is easy to see that the integration over
$d^2zd^2w$ produces a real result, and thus the entire contribution
from the even spin structure vanishes.

Indeed, first observe that $e^{-<k\cdot X(z) p\cdot X(w)>}$ is real,
then also that each term in the above expression,
call it $T(z,w)$, is even by
doing simultaneously
$z\to -z,\bar z\to -\bar z,w\to -w,\bar w\to -\bar w$.
Thus
\beqa 
 \int^{1/2}_{-1/2}dRez\int^{1/2}_{-1/2}dRew T(z,w)=
\int^{1/2}_{-1/2}dRez\int^{1/2}_{-1/2}dRew T(\bar z,\bar w)
\eeqa
i.e. the result is real.

Also notice that the first four terms do not contain any fermionic 
propagators and vanish in any case because of the sum over spin structures.

\vspace*{0.5cm}
\appendix{\bf Appendix D. Analysis of the field theory limit} 

\renewcommand{\theequation}{D.\arabic{equation}}
\setcounter{equation}{0}

In this Appendix we analyze the small $q^2$ limit of the amplitude,
arguing that the leading behaviour comes from the pinching limit
$z\sim w$
of the term proportional to  
$\vert F(z-w)\vert ^2$ in eq. (\ref{final}). This leading result, which
provides the strongest singularity for $q^2\to 0$ and thus the
leading behaviour at large distances, has been analyzed in Section 
\ref{pinch}.
Here we look for the $q^2\rightarrow 0$ limit of every term of eq. 
(\ref{final}), 
including $\vert F(z-w)\vert ^2$ without taking the pinching limit.
The leading behaviour comes from the region of the 
moduli space $l\to\infty$ and $Imz,Imw\sim l$.

First we analyze the terms of the form $F(x_1-x_2)F(x'_1-x'_2)$, leaving
aside those of the form 
$\partial_{x_1}\partial_{x_2} P_D$ for a later discussion.

Let us define $\zeta = {Im z\over l}$ and $\eta ={Im w\over l}$ and use the
following
 asymptotic behaviour for $l\rightarrow \infty$
\beq
{1\over \pi} F(z-w) \rightarrow {cos \pi (z-w) \over {sin \pi (z-w)}} + i 
(\zeta - \eta)
\eeq
Thus, the limit $l \rightarrow \infty$ at fixed $\zeta ,~\eta$  
of the different terms appearing in the amplitude is 
\beqa
{1\over \pi}F(z-w) \rightarrow i[-\epsilon (\zeta-\eta) + \zeta - \eta]\nn\\
{1\over \pi}F(\bar z-\bar w) \rightarrow -i
[-\epsilon (\zeta-\eta) + \zeta - \eta]\nn\\
{1\over \pi}F(z-\bar w) \rightarrow i[- 1 + \zeta + \eta]\nn\\
{1\over \pi}F(\bar z-w) \rightarrow i[- 1 + \zeta + \eta]\nn\\
{1\over \pi}F(w-\bar w) \rightarrow i[- 1 + 2 \eta]
\eeqa
(where $\epsilon (\pm x)=\pm 1$) and moreover
\beqa
 e^{-<k\cdot X(z) p\cdot X(w)>}\to 
e^{-2\pi q^2l\eta (1-\zeta )+2\pi k_0^2l(\eta -\zeta )^2}
~~for~~\zeta >\eta \nn\\
e^{-2\pi q^2l\zeta (1-\eta )+2\pi k_0^2l(\eta -\zeta )^2}
~~for~~\eta >\zeta 
\eeqa

In every term we first do the integration on the difference
$(\eta -\zeta )$ taking advantage of the gaussian factor
$e^{2\pi k_0^2l(\eta -\zeta )^2}$, interpreting it as the
analytic continuation from $k_0^2<0$, obtaining a factor
${1\over \sqrt{-k_0^2}\sqrt{l}}$ times the rest of the
integrand for $\eta =\zeta$.

We have then for the terms proportional to 
$aq^2(q\cdot h\cdot q)$ in eq. (\ref{final}) the integral 
(keeping the large $l$ contribution) 
\beqa
\int dll^{-3/2}{l^2 \over \sqrt{-k_0^2}\sqrt{l}} \int_0^1
d\zeta\zeta (1- \zeta)\cdot e^{-2\pi q^2l\zeta (1-\zeta )}
\eeqa
both for 
$\zeta >\eta$ and for $\eta >\zeta$. From this last integral
we get a behaviour $1/q^2\sqrt{-k_0^2}$, which gives for 
the amplitude $aq\cdot h\cdot q/\sqrt{-k_0^2}$, i.e. subleading
with respect to the behaviour coming from the pinching limit.

Let us consider next the terms in eq. (\ref{final}) proportional to
$aq\cdot h\cdot q~k_0^2$. It is seen from the above equations
on the $l\to\infty$ behaviour that altogether they give
an integrand of the form
\beqa
f(\zeta ,\eta )\cdot  e^{-<k\cdot X(z) p\cdot X(w)>}
\eeqa
where $f(\zeta =\eta )=0$. From that one can expect, as it is confirmed by
a more detailed analysis, that the subsequent integrations
will give a result for the amplitude not worse than  
$aq\cdot h\cdot qlog(q^2)/\sqrt{-k_0^2}$ for $q^2\to 0$.

Notice that actually the set of terms
\beqa
-F(z-\bar w)F(\bar z-\bar w)-F(z-w)F(\bar z-w)
+F(z-w)F(z-\bar w)+F(\bar z-\bar w)F(\bar z-w)
\eeqa
vanishes upon the symmetric integration over $d^2zd^2w$.

Finally let us consider the terms of the kind 
$\partial_{x_1}\partial_{x_2} P_D$, where 
$x_1$ can be $z$ or $\bar z$ etc. Consider for instance the term 
$x_1=z,~x_2=w$ (the discussion for the others being similar). 
Its behaviour for $l\to\infty$ is
\beqa
\sim {1 \over sin^2\pi (z-w)}+{1 \over l}
\eeqa
We then take $z=Rez+il\zeta$ ,$w=Rew+il\eta$ and the asymptotic
behaviour of $1/sin^2$ gives a vanishing contribution due
to the integration over $dRezdRew$. There remains the term 
${1\over l} e^{-<k\cdot X(z) p\cdot X(w)>}$, which, after the gaussian 
integration described above and the subsequent integrations,
gives for the amplitude again a result like
$aq\cdot h\cdot qlog(q^2)/\sqrt{-k_0^2}$ for $q^2\to 0$.

\end{document}